\documentclass[aps,pre,showpacs,superscriptaddress,twocolumn]{revtex4}

\usepackage[british]{babel}
\usepackage{graphicx}
\usepackage{amsmath}
\usepackage{amssymb}
\usepackage{mathrsfs}
\usepackage{color}
\usepackage{hyperref}

\definecolor{gnu4}{RGB}{255,165,0}

\newcommand{\FH}[7]{H^{#1#2}_{#3#4}\!\!\left[ #5 \left| \begin{array}{l l} #6 
\vspace{2pt}\\ #7 \end{array} \right. \right]}
\newcommand{\Lt}[3]{\mathscr{L}_{#2\rightarrow #3}\!\left\{#1\right\}}
\newcommand{\Lti}[3]{\mathscr{L}^{-1}_{#2\rightarrow #3}\!\left\{ #1 \right\}}
\newcommand{\ibe}[3]{B\!\left(#1;\;#2,#3\right)}
\newcommand{\gmle}[3]{E_{#1,#2}\!\left(#3\right)}
\newcommand{\mle}[2]{E_{#1}\!\left(#2\right)}

\begin{document}

\title{Aging renewal theory and application to random walks}

\author{Johannes H. P. Schulz}
\affiliation{Physics Department T30g, Technical University of Munich,
85747 Garching, Germany}
\author{Eli Barkai}
\affiliation{Department of Physics, Bar Ilan University, Ramat-Gan 52900,
Israel}
\author{Ralf Metzler}
\affiliation{Institute for Physics \& Astronomy, University of Potsdam,
14476 Potsdam-Golm, Germany}
\affiliation{Physics Department, Tampere University of Technology,
FI-33101 Tampere, Finland}

\date{\today}

\begin{abstract}
The versatility of renewal theory is owed to its abstract formulation.
Renewals can be interpreted as steps of a random walk, switching events
in two-state models, domain crossings of a random motion, etc.
We here discuss a renewal process in which successive events are separated
by scale-free waiting time periods. Among other ubiquitous long time
properties, this process exhibits  aging: events counted initially in a
time interval $[0,t]$ statistically strongly differ from those observed at
later times $[t_a,t_a+t]$.
In complex, disordered media, processes with scale-free waiting times
play a particularly prominent role. We set up a unified analytical foundation
for such anomalous dynamics by discussing in detail the distribution of
the aging renewal process. We analyze its half-discrete, half-continuous
nature and study its aging time evolution. These results are readily used
to discuss a scale-free anomalous diffusion process, the continuous time
random walk. By this we not only shed light on the profound origins of its
characteristic features, such as weak ergodicity breaking. Along the way,
we also add an extended discussion on aging effects. In particular, we find
that the aging behavior of time and ensemble averages is conceptually very
distinct, but their time scaling is identical at high ages. Finally, we show
how more complex motion models are readily constructed on the basis of aging
renewal dynamics.
\end{abstract}


\maketitle

\section{Introduction}
A stochastic process $n(t)$ counting the number of some sort of events occurring during a time interval $[0,t]$ is called a renewal process, if the time spans between consecutive events are independent, identically distributed random variables~\cite{Cox}. Renewal theory does not specify the exact meaning or effect of a single event. It could be interpreted as the appearance of a head in a coin tossing game, the arrival of a bus or of a new customer in a queue. In a mathematical formulation, events remain abstract objects characterized by the time of their occurrence. Thus, not surprisingly, renewal processes are at the core of many stochastic problems found throughout all fields of science. Maybe the most obvious physical application is the counting of decays from a radioactive substance. This is an example of a Poissonian renewal process: the random time passing between consecutive decay events, the waiting time, has an exponential probability density function $\psi(t)=\tau^{-1} \exp(-t/\tau)$. In other words, events here are observed at a constant rate $\tau^{-1}$ (that is, if the sample is sufficiently large and the half-life of individual atoms sufficiently long).

A physical problem of more contemporary interest is subrecoil laser cooling~\cite{Bardou,Lutz2004}. Two counterpropagating electromagnetic waves can cool down individual atoms to an extent where they randomly switch between a trapped (i.e. almost zero momentum) state and a photon emitting state. Successive life times of individual states are found to be independent and stationary in distribution. Hence, the transitions from the trapped to the light emitting state form the events of a renewal process. Similar in spirit, colloidal quantum dots \cite{Brokmann2003,Margolin2004} switch between bright states and dark states under continuous excitation. In contrast to the Poissonian decay process, the latter two examples feature a power-law distribution of occupation times $t$, whose long time asymptotics reads
\begin{equation}\label{intro:tail}
  \psi(t)\sim \frac{\tau^\alpha}{|\Gamma(-\alpha)|\,t^{1+\alpha}} 
\qquad\text{with}\qquad 0<\alpha<1.
\end{equation}
Such heavy-tailed distributions are not uncommon for physically relevant renewal processes. To see this, consider a simple unbounded, one-dimensional Brownian motion, and let $n(t)$ count the number of times the particle crosses the origin. Then the waiting time between two crossings is of the form~(\ref{intro:tail}) with $\alpha=1/2$. Indeed, a random walk of electron-hole pairs either in physical space, or in energy space, was proposed as a mechanism leading to the power-law statistics of quantum dot blinking \cite{Shimizu2001}. In general, whenever events are triggered by domain crossings of a more complex, unbounded process, power-law distributed waiting times are to be expected. In addition, the latter can be interpreted as a superposition of exponential transition times with an (infinitely) wide range of rate constants $\tau$. The power-law form (\ref{intro:tail}) for the inter-event statistics is a typical ansatz to explain such renewal dynamics in highly disordered or heterogeneous media such as spin glasses~\cite{Bouchaud1992}, amorphous semiconductors~\cite{Scher1975} or biological cells~\cite{cells}.

Note that distributions as in~(\ref{intro:tail}) imply a divergent average waiting time, $\langle t\rangle=\int_0^\infty t\,\psi(t)\,\mathrm{d}t=\infty$. Renewal processes of this type are said to be scale-free, since, roughly speaking, statistically dominant waiting times are always of the order of the observational time. Hence, their outstanding characteristics play out most severely on long time scales: while for $t\gg\tau$, Poissonian renewal processes behave quasi-deterministically, $n(t)\approx t/\tau$, heavy-tailed distributions lead to nontrivial random properties at all times. Stochastic processes of this type are known to exhibit weak ergodicity breaking~\cite{Rebenshtok2008}, i.e. time averages and associated ensemble averages of a physical observable are not equivalent. Moreover, despite the renewal property, the process $n(t)$ is nonstationary~\cite{Godreche2001}: events $n_a(t_a,t)$ counted after an unattended aging period $t_a>0$, i.e. within some time window $[t_a,t_a+t]$, are found to be statistically very distinct from countings during the initial period $[0,t]$. Fewer events are counted during late measurements, in a statistical sense, and thus we also say the process exhibits \textit{aging}. Deeper analysis reveals that this slowing down of dynamics is due to an increasingly large probability to count no events at all during observation. Intuitively, in the limit of long times $t,t_a\gg\tau$, one would expect to see at least some renewal activity $n_a>0$. Instead, the probability to have exactly $n_a=0$ increases steadily, and as $t_a\rightarrow\infty$, the system becomes completely trapped.

The first part of our article is devoted to an in-depth discussion of the aging renewal process. We directly build on the original work of~\cite{Barkai2003a,Barkai2003b,Godreche2001}, calculating and discussing extensively the PDF of the aged counting process $n_a$, with a special emphasis on aging mediated effects. We provide leading order approximations for slightly and strongly aged systems, and discuss their validity. We then derive the consequences for related ensemble averages, and discuss the special impact of the uneventful realizations.

The second part addresses the question of how these peculiar statistical effects translate to physical applications of renewal theory. Our case study is the celebrated continuous time random walk (CTRW) \cite{Hughes}.
Originally introduced to model charge carrier transport in amorphous semiconductors~\cite{Scher1975}, it has been successfully applied to many physical and geological problems~\cite{report}, and was identified as diffusion process in living cells~\cite{cells}. It extends standard random walk models due to the possibility to have random waiting times between consecutive steps. Heavy-tailed waiting times~(\ref{intro:tail}) are of course of special interest to us, and we review several well studied phenomena inherent to CTRW models in the light of aging renewal theory. Indeed, the insights gained from the first part of our discussion help us to contribute several new aspects to the bigger picture, especially with respect to aging phenomena (see also~\cite{Barkai2003a,Barkai2003b,Monthus1996,Schulz2013}). Thus we analyze the scatter of time averages and their deviation from the corresponding ensemble averages (the fingerprint of weak ergodicity breaking \cite{He2008,Lubelski2008,Neusius2009,Deng2012,Sokolov2009,Lomholt2007,Burov2010}) under aging conditions. We report strong conceptual differences between aging effects on time and ensemble averages, yet they exhibit identical time scalings at late ages. Furthermore, we discuss a novel population splitting mechanism, which is a direct consequence of the discrete, increasing probability to measure $n_a=0$ steps during late observation: a certain fraction within an ensemble of CTRW particles stands out statistically from the rest as being fully immobilized. Finally, we use the methods and formulae developed in the course of our discussion to analyze a model combination of CTRW and fractional Langevin equation motion~\cite{Pottier2003}. By this we include the effects of binding and friction forces, and a persistent memory component. We discuss the intricate interplay of aging and relaxation modes and highlight the essential features and pitfalls for aged ensemble and single trajectory measurements.

\section{Aging renewal theory}\label{renewal}
We here analyze in detail a renewal process with distinct non-Markovian characteristics, focusing especially on its aging properties. In that, we proceed as follows. In section~\ref{renewal:process}, we define the aging renewal process and introduce some basic notation and concepts to be used throughout the rest of this work. We then turn to a long time scaling limit description of the renewal process in section~\ref{renewal:scalinglimit}. In the case of scale-free waiting times, we obtain a continuous time counting process with interesting nonstationary random properties. Sections~\ref{renewal:process} and~\ref{renewal:scalinglimit} have review character, and we refer to the original works for more detailed discussions on generalized limit theorems. Here, we concentrate on the time evolution of the renewal probability distribution, which we study extensively in section~\ref{renewal:distribution}. In particular, we discuss the emergence of both a discrete and a continuous contribution and contrast slightly and highly aged systems in terms of the behavior around the origin, around local maxima, and in the tails. The implications for calculating and interpreting ensemble averages are deduced in section~\ref{renewal:averages}.  With the probability distribution separating into discrete and continuous contributions, it is natural to consider conditional averages, an issue we address in section~\ref{renewal:conditionalaverages}.

\begin{figure*}
 \includegraphics{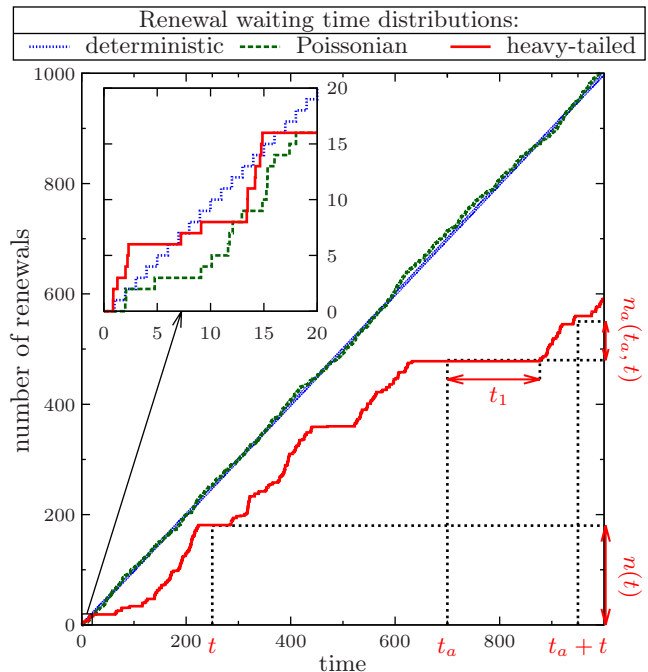}
 \caption{Sample realizations for three different types of renewal processes. 
Events are separated by waiting times, which are independent, identically 
distributed according to a probability density function (PDF) $\psi(t)$. We 
depict here the case of deterministic periodic renewals, $\psi(t)=\delta(t-1)$, 
of Poissonian waiting times, $\psi(t)=e^{-t}$, and of heavy-tailed waiting 
times, $\psi(t)=4.5\cdot(5t+1)^{-1.5}$, see key. \textit{Inset:} During an 
observation on short time scales, the randomness of Poissonian and heavy-tailed 
waiting times contrasts the regular, steady progression of the deterministic 
renewal process. However, it is difficult to distinguish the two random 
processes on this level of analysis. \textit{Main figure:} Observation on long
time scales reveals the profound statistical difference between the two random 
processes: The Poissonian renewal process behaves almost deterministically on 
time scales long as compared to $\langle t \rangle=1$. At the same time, the 
random nature of heavy-tailed waiting times is visible on all time scales, as 
$\langle t \rangle=\infty$. On the one hand, this means that counting the number 
of renewal events $n$ up to time $t$ yields different results from one process 
realization to the next: $n(t)$ remains a nontrivial random counting process on 
all time scales. On the other hand, for ensemble measurements aging should be 
taken into account: The statistics of the number of renewals $n(t)$ during 
observation time $[0,t]$ can turn out to be different from the statistics of 
$n_a(t_a,t)$, the number of renewals in $[t_a,t_a+t]$. The  conceptual 
difference of the two counting processes lies in the forward recurrence time 
$t_1$, which measures the time span between the start of observation at time 
$t_a$ and the counting of the very first event, see graph. If the observation 
starts simultaneously with the renewal process, i.e. $t_a=0$ and $n_a\equiv n$, then $t_1$ 
has the PDF $\psi(t_1)$, just like any other waiting time. But when $t_a>0$,
then $t_1$ typically
only represents the observed fraction of a regular waiting time, which possibly 
started before $t_a$. Therefore, $t_1$ has its own distinct, age-dependent PDF 
$h(t_a,t_1)$.
 }
 \label{fig.renprocess}
\end{figure*}

\subsection{The aging renewal process}\label{renewal:process}
Suppose we are interested in a series of events that occur starting from time $t=0$.
We may later choose to specifically identify these events with the arrival of a 
bus, the steps of a random walk, or the blinking of a quantum dot---for now, 
they remain abstract. Let $n(t)$ count the number of events that occurred up to 
time $t$; we occasionally refer to it as a counting process. The time spans 
between two consecutive events are called waiting times. They are not 
necessarily fixed. Instead, we take them to be independent, identically 
distributed random variables. In such a case, it is justified to refer to events 
as \emph{renewals}: The process $n(t)$ is not necessarily Markovian, but any 
memory on the past is erased with the occurrence of an event---the process is 
renewed. The probability density function (PDF) of individual waiting times is
denoted by~$\psi(t)$. Obviously, the nature of this quantity heavily 
influences the statistics of the overall renewal process. 
Figure~\ref{fig.renprocess} shows realizations for deterministic periodic 
renewals, $\psi(t)=\delta(t-\tau)$, for Poissonian waiting times, 
$\psi(t)=\tau^{-1}\exp(-t/\tau)$, and for heavy-tailed waiting times, 
i.e. $\psi(t)$ has the long-$t$ asymptotics~(\ref{intro:tail}). In all cases, the scaling parameter $\tau>0$ serves as a microscopic time scale for individual waiting times.

First, study the inset of Fig.~\ref{fig.renprocess}, which focuses on the 
initial evolution of these processes at short time scales. The complete
regularity of the deterministic renewals is distinct, but the two random
counting processes are not 
clearly discernible by study of such single, short-period observations. Now 
compare this to the main figure, which depicts realizations of the processes on
much longer time scales. Here, the realizations of the deterministic and 
Poissonian renewal processes look almost identical. Recall that for independent 
exponential waiting times, the average time elapsing until the $n$th step is 
made increases linearly with $n$, while the fluctuations around this average 
grow like $n^{1/2}$. Thus, the relative deviation from the average decays to 
zero on longer scales. Roughly speaking, on time scales that are long as 
compared to the average waiting time $\langle t\rangle=\tau$ we observe a quasi-deterministic relation 
$n(t)=t/\tau$.

For heavy-tailed waiting times as in~(\ref{intro:tail}), the picture is 
inherently different. Above scaling arguments fail, since the typical time scale 
to compare to, the average $\langle t\rangle$ of a single waiting time, is 
infinite. This is why this type of dynamics are sometimes referred to as 
scale-free and they are studied in light of generalized central limit theorems 
\cite{gclt}. We sketch some of the analytical aspects in the following 
section. Most importantly, it turns out that in the absence of a typical time 
scale, waiting time periods persist and are statistically relevant on 
arbitrarily long time scales. The effect is clearly visible in 
Fig.~\ref{fig.renprocess}: The renewal process $n(t)$ remains a nontrivial 
random process, even when $t\gg\tau$.

We also introduce at this point the concept of an aged measurement: while the 
renewal process starts at time $0$, an observer might only be willing to or 
capable of counting events starting from a later time $t_a>0$. In place of the 
total number of renewals $n(t)$ he then studies the counted fraction 
$n_a(t_a,t)=n(t_a+t)-n(t_a)$. The fundamental statistical difference between the 
renewal processes $n$ and $n_a$ stems from the statistics of the time period 
$t_1$ which passes between start of the measurement at $t_a$ and the observation 
of the very first event. We refer to it as the \textit{forward recurrence time} 
\cite{Godreche2001} and denote its PDF by $h(t_a,t_1)$. If the observer counts starting  at time $t_a=0$, the forward recurrence time is simply distributed like any other waiting time, $h(0,t_1)=\psi(t_1)$. But for later, aged measurements, 
$t_a>0$, the distribution is different, as indicated in Fig.~\ref{fig.renprocess}.

We call the dependence of the statistical properties of the counted renewals 
$n_a$ on the starting time of the measurement $t_a$ an aging effect. Its impact crucially depends on the waiting time distribution in use. For instance, a Poissonian renewal process is a Markov process, meaning here that events at all times occur at a constant rate. In this case, $h(t_a,t)\equiv\psi(t)$, so there the process does not age. For any other distribution, $h(t_a,t)\neq\psi(t)$; yet, if the average waiting time is finite, then on time scales long in comparison to the average
waiting time $\langle t\rangle$, the renewal process behaves quasi-deterministically (details below). Thus aging is in effect, but becomes negligible at long times. But 
scale-free waiting times as in~(\ref{intro:tail}) result in nontrivial 
renewal dynamics, with distinct random properties, and aging effects 
should be taken into careful consideration. In the following section, we thus 
study and compare in detail the statistics of the renewal processes $n(t)$ and 
$n_a(t_a,t)$ in terms of their probability distributions and discuss the 
consequences for calculating aged ensemble averages.

\subsection{Long time scaling limit}\label{renewal:scalinglimit}
Several authors have studied the aging renewal process as defined above and its long time approximation, see \cite{Godreche2001,Bingham1971,Meerschaert2004,Hughes} and references therein. We demonstrate here the basic concept of a scaling limit, focusing on the calculation of the rescaled PDF.

The probability $p(n;t)$ of the random number of events $n$ taking place up to 
time $t$ takes a simple product form in Laplace space \cite{footnote:laplacetrafo},
\begin{align}
  p(n;s) &= \Lt{p(n;t)}{t}{s} = \int_0^\infty e^{-st} p(n;t) \,dt \nonumber\\
  &= \psi(s)^n  \, \frac{1-\psi(s)}{s} ,
 \label{renewal:pdfL_psi}
\end{align}
which is a direct consequence of the renewal property of the process. It can be 
read as the probability to count exactly $n$ steps at some arbitrary 
intermediate points in time and not seeing an event ever since, i.e. the
expression $[1-\psi(s)]/s$ is the Laplace transform of $1-\int_0^t\psi(t)dt$.

We also seek to study measurements taken after some time period $t_a$ during 
which the process evolved unattendedly. To this intent, we should consider 
$n_a(t_a,t)=n(t_a+t)-n(t_a)$, the number of events that happen during the time 
interval $[t_a,t_a+t]$. The corresponding probability $p_a(n_a;t_a,t)$ reads in 
double Laplace space, $(t_a,t)\rightarrow(s_a,s)$, \cite{Godreche2001}
\begin{subequations}
\label{renewal:aPDFL_psiA}
\begin{equation}
p_a(n_a;s_a,s)=\left\{\begin{array}{ll}
(s_as)^{-1}-h(s_a,s)s^{-1}, & n_a=0\\[0.2cm]
\displaystyle h(s_a,s)\psi^{n_a-1}(s)\frac{1-\psi(s)}{s}, & n_a\geq1
\end{array}\right.,
\label{renewal:aPDFL_psi}
\end{equation}
where we introduced
\begin{equation}
h(s_a,s)=\frac{\psi(s_a)-\psi(s)}{s-s_a}\frac{1}{1-\psi(s_a)},
\label{renewal:recL_psi}
\end{equation}
\end{subequations}
the PDF of the forward recurrence time $t_1$ as defined
in the preceding section. The interpretation of Eq.~(\ref{renewal:aPDFL_psi}) is 
straightforward: The probability to see any events at all during the period of 
observation equals the probability that $t_1\leq t$. Furthermore, the observer 
counts exactly $n_a$ events if the first event at time $t_a+t_1$ is followed by 
$(n_a-1)$ events at intermediate times $t_a+t_i$ and an uneventful time period 
until the measurement ends at $t_a+t$.

We check that for Poissonian waiting times, $\psi(t)=\tau^{-1}\exp(-t/\tau)$, we have $\psi(s)=1/(s\tau+1)=s_a h(s_a,s)$, and hence $h(t_a,t)\equiv\psi(t)$. The Poissonian (i.e. Markovian) renewal process is unique in this respect. We say, it does not age.

Now assume that waiting times are heavy-tailed, i.e. their PDF is of the 
form~(\ref{intro:tail}). Waiting times of this type have a diverging mean 
value, which has severe consequences for the resulting renewal process, even in 
the scaling limit of long times. To see this, introduce a scaling constant 
$c>0$ and rescale time as $t\mapsto t/c$. Then write the following approximation 
in Laplace space, where, by virtue of Tauberian theorems \cite{Hughes}, small Laplace variables correspond to long times,
\begin{equation}
  \Lt{c\psi(ct)}{t}{s} = \psi(s/c) \approx 1-(s\tau/c)^\alpha .
  \label{renewal:tailL}
\end{equation}
If we rescale the counting process accordingly, meaning here $n\mapsto 
n/c^{\alpha}$, we can take the limit $c\rightarrow\infty$ to arrive at a long
time limiting version of the renewal process. For instance, the probability 
distribution in Eq.~(\ref{renewal:pdfL_psi}) takes the following form:
\begin{subequations}
\label{renewal:pdfLA}
\label{renewal:pdf}
\begin{align}
  p(n;s) &\mapsto \Lt{c^{\alpha}p(nc^\alpha;ct)}{t}{s} = 
c^{\alpha-1}p(nc^\alpha;s/c) \nonumber\\
  &\approx \tau^\alpha s^{\alpha-1} \left[ 1-\frac{n(s\tau)^\alpha}{nc^\alpha}\right]^{nc^\alpha}  \nonumber\\
&\rightarrow \tau^\alpha s^{\alpha-1} \exp[-n(s\tau)^\alpha] \quad (c\rightarrow\infty) .
\label{renewal:pdfL}
\end{align}
Note that in these equations, $n$ was turned from an integer to a continuous 
variable, characterized by a PDF. Still, for simplicity, we continue to 
refer to this variable as `number of events'. For the sake of notational convenience we set $\tau=1$ in what follows, bearing in mind that the rescaled time variables $t$ and $t_a$ are measured in units which are by definition large as compared to the microscopic time scale of individual waiting times, the scaling factor $\tau$ in Eq.~(\ref{intro:tail}).

\begin{figure*}
 \includegraphics[width=16cm]{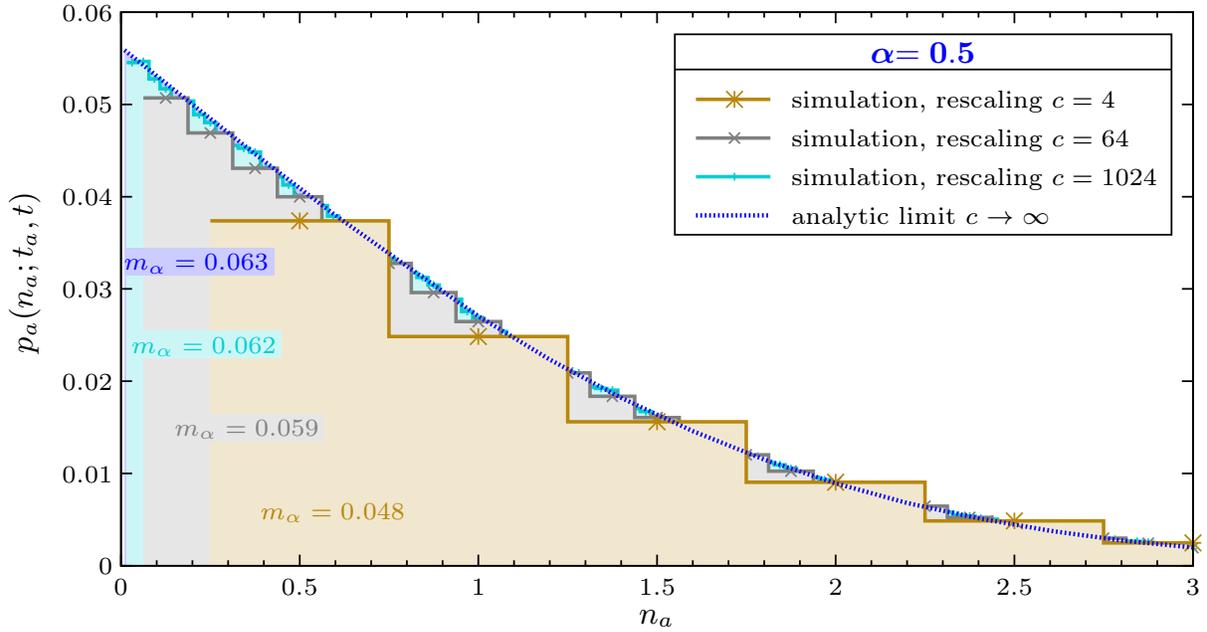}
 \caption{Scaling convergence of the aging renewal process. We simulate a renewal process with a waiting time distribution $\psi(t)=(4\pi)^{-1/2}(\pi^{-1}+t)^{-3/2}$. The latter is of the heavy-tailed form~(\ref{intro:tail}) with $\alpha=0.5$ and $\tau=1$ (a.u.). Time is rescaled as $t\mapsto t/c$ and renewals as $n_a\mapsto n_a/c^{\alpha}$. We plot the PDF $p_a(n_a;t_a,t)$ for the number of renewals within the time interval $[t_a,t_a+t]$, with $t_a=100$ and $t=1$, in terms of the rescaled quantities. The area below each step in the graph represents the probability to count a certain number of (rescaled) renewals; the respective values are indicated by symbols. As the scaling constant $c$ increases (see key), $p_a$ converges to the analytical, smooth scaling limit, i.e. the PDF given through Eqs.~(\ref{renewal:aPDFA}) [same as in Fig.~\ref{fig.PDF}, right center panel]. Since here, we only show renewal probabilities for $n_a>0$, the total area below each graph equals the probability $m_\alpha$ to count any events at all. Ensemble statistics are based on data from $10^7$ independent renewal process realizations.
 }
 \label{fig.hist}
\end{figure*}

The evolution of the probability density with respect to real time $t$ is now given through
\begin{equation}
p(n;t)=\frac{1}{\alpha}tn^{-1-1/\alpha} \ell_\alpha^+(nt^{-\alpha}),
\end{equation}
where
\begin{equation}
\ell_\alpha^+(z)=\Lti{\exp(-s^\alpha)}{s}{z}.
\end{equation}
\end{subequations}
Remarkably, $n(t)$ thus remains a nontrivial random quantity even after the rescaling procedure. The special limit $\alpha\rightarrow1$ is representative for finite average waiting times. In this case, the Laplace transform in Eq.~(\ref{renewal:tailL}) is a moment generating function, and thus we identify $\tau=\langle t\rangle$. In the scaling limit such a process collapses to a deterministic counting process: Eqs.~\eqref{renewal:pdf} then imply $p(n;t) = \delta(n-t)$.

In contrast, for any $0<\alpha<1$, $n(t)$ obeys a scaling relation $n(t)\sim t^\alpha$ and follows a 
Mittag-Leffler law \cite{Bingham1971}, directly related to the one-sided stable density $\ell_\alpha^+(z)$ \cite{gclt,footnote:plotstable}. The latter is a fully continuous function on the positive half-line $z\geq0$. This implies in particular that for $t>0$, the probability to have exactly $n(t)=0$ is 
infinitely small. Apparently, for the long time scaling limit of the counting 
process $n(t)$, the length of the very first single waiting time is negligible 
and the observer starts counting events immediately after initiation of the 
process.

The procedure for finding the PDF $p_a$ of counted events $n_a$ in an aged 
measurement, $t_a\geq0$, is analogous. In the long time scaling limit defined 
above, Eqs.~(\ref{renewal:aPDFL_psiA}) turn into
\begin{subequations}
\begin{eqnarray}
\nonumber
p_a(n_a;s_a,s)&=&\delta(n_a)\left[\frac{1}{s_as}-m_\alpha(s_a,s)\right]\\
&&+h(s_a,s)p(n_a;s)\label{renewal:aPDFL},
\end{eqnarray}
with the definition
\begin{equation}
m_\alpha(s_a,s)=h(s_a,s)/s,
\end{equation}
and
\begin{equation}
h(s_a,s)=\frac{s_a^\alpha-s^\alpha}{s_a^\alpha(s_a-s)}.
\label{renewal:recL} 
\end{equation}
\label{renewal:recLA}
\end{subequations}
Eq.~\eqref{renewal:aPDFL} demonstrates the aging time's distinct influence on 
the shape of the PDF of the number of events. Most remarkably, as $t,t_a>0$,
the occurrence of a term proportional to $\delta(n_a)$ indicates a nonzero 
probability for counting exactly $n_a(t_a,t)=0$. This means that we might observe no events 
at all in the time interval $[t_a,t_a+t]$. This is a quite distinct aging 
effect, contrasting the immediate increase of the non-aged counting $n(t)$. Only 
the limit $\alpha\rightarrow1$ leads us back to a trivial deterministic, 
nonaging counting process, and consequently $p_a(n_a;t_a,t)\equiv p(n_a;t) = 
\delta(n_a-t)$.

For the aged PDF $p_a$, Laplace inversion of Eqs.~(\ref{renewal:pdfLA}) and
(\ref{renewal:recLA}) to real time $t_a,t$ yields \cite{Barkai2003a,Barkai2003b}
\begin{subequations}
\label{renewal:aPDFA}
\begin{eqnarray}
p_a(n_a;t_a,t)&=&\delta(n_a)\left[1-m_\alpha(t_a,t)\right]\nonumber\\
&&+h(t_a,t)*_tp(n_a;t) \label{renewal:aPDF}
\end{eqnarray}
with \cite{Godreche2001,Dynkin1961,Feller}
\begin{eqnarray}
m_\alpha(t_a,t)&=&\int_0^t h(t_a,t')\mathrm{d}t'\nonumber\\
&=&\frac{\ibe{[1+t_a/t]^{-1}}{1-\alpha}{\alpha}}{\Gamma(1-\alpha)\Gamma(\alpha)} 
\nonumber\\
&&\equiv m_\alpha(t_a/t),\label{renewal:malpha}
\end{eqnarray}
where
\begin{equation}
h(t_a,t)=\frac{\sin(\pi\alpha)}{\pi}\frac{t_a^\alpha}{t^\alpha(t_a+t)} 
\label{renewal:rec}.
\end{equation}
\end{subequations}
Here the asterisk $*_t$ indicates a Laplace convolution with respect to time $t$. Figure~\ref{fig.hist} gives a first example of how such an aged PDF behaves. We depict the case $\alpha=0.5$ at high ages, $t_a/t=100$. In addition we demonstrate scaling convergence: if a renewal process with simple power-law waiting time distribution is monitored on increasingly long scales for time and event numbers, then its statistics approach the continuous limit described by Eqs.~(\ref{renewal:aPDFA}).

The latter relate the aged PDF $p_a$ to the non-aged PDF $p$ via the PDF 
$h$ of the forward recurrence time. $m_\alpha$ is the probability to count any 
events at all during observation. Its representation in terms of an incomplete 
Beta function $\ibe{z}{a}{b}$ (see Appendix \ref{app:functions}) is found by a 
simple substitution $u=t'/(t'+t_a)$. It can be written as a function of the 
ratio $t_a/t$ alone and we suggest to use the latter as a more precise and quantitative notion of the \textit{age} of the measurement. In particular, we call the process or measurement or observation \textit{slightly aged}, if $t_a\ll t$. Conversely, we say it is \textit{strongly} or \textit{highly aged}, if $t_a\gg t$. We now look deeper into these two limiting regimes.

\subsection{Aging probability distribution}\label{renewal:distribution}

\textit{Slightly aged PDF. ---}
First, notice that in Eqs.~(\ref{renewal:aPDFA}), the PDF $h$ 
of the forward recurrence time appears inside integrals,
and should be interpreted in a distributional sense. For 
instance, in the limit $t_a\rightarrow0$ we should recover the PDF of a non-aged 
system, $p_a\rightarrow p$. To confirm this, study the limit 
$s_a\rightarrow\infty$ in Laplace space. We find $h(s_a,s)\sim s_a^{-1}$. Thus 
we should write $h(0,t)=\delta(t)$ in terms of a Dirac $\delta$-distribution, 
consistently implying $m_\alpha(0)=1$ and $p_a(n_a;0,t)\equiv p(n_a;t)$. Again 
we find that only an observer counting from the initiation of the (rescaled) 
renewal process, $t_a=0$, witnesses the onset of activity instantly.

We can go one step beyond this limit approximation and study the properties of a 
slightly aged system ($t_a\ll t$ and $s_a\gg s$). We write $h(s_a,s)\sim 
s_a^{-1}-s_a^{-1-\alpha}s^\alpha$ and find, by use of Tauberian theorems that
\begin{equation}
1-m_\alpha(t/t_a)\sim\frac{(t_a/t)^\alpha}{\Gamma(1+\alpha)\Gamma(1-\alpha)},
\end{equation}
and
\begin{subequations}
\label{renewal:aPDF_early_LaplaceA}
\begin{eqnarray}
\nonumber
h(t_a,t)*_t p(n_a;t)&\sim&\Lti{s^{\alpha-1}e^{-n_a s^\alpha}}{s}{t}\\
\nonumber
&&-\frac{(t_a/t)^\alpha}{\Gamma(1+\alpha)}\Lti{s^{2\alpha-1}e^{-n_as^\alpha}}
{s}{t}\\
&&\hspace*{-2.4cm}
\sim p(n_a;t)-\frac{t_a^\alpha}{\Gamma(1+\alpha)}\Lti{s^{2\alpha-1}e^{-n_as
^\alpha}}{s}{t}.
\label{renewal:aPDF_early_Laplace}
\end{eqnarray}
Here, first order corrections are provided in terms of a Laplace inversion. For the analytical discussion, we can alternatively express them as
\begin{equation}
 h(t_a,t)*_t p(n_a;t) \sim p(n_a;t) + \frac{t_a^\alpha}{\Gamma(1+\alpha)} \frac{\partial p(n_a;t)}{\partial n_a} ,
 \label{renewal:aPDF_early_invstable}
\end{equation}
relating them to the non-aged PDF, and thus to the familiar stable density, see Eqs.~(\ref{renewal:pdf}). Finally, we can also interpret these contributions as Fox-$H$ functions, for which we know series expansions for small arguments and asymptotics for large arguments (see Appendix \ref{app:functions} and~\cite{Mathai}; the connection between Fox-$H$ functions, stable densities and fractional calculus is established in~\cite{Schneider}):
\begin{eqnarray}
\nonumber
h(t_a,t)*_t p(n_a;t)&\sim&\FH{1}{0}{1}{1}{\frac{n}{t^\alpha}}{(1-\alpha,\alpha)}
{(0,1)}\\
&&\hspace*{-2cm}
-\frac{(t_a/t)^\alpha}{\Gamma(1+\alpha)}\frac{1}{t^\alpha}\FH{1}{0}{1}{1}{
\frac{n}{t^\alpha}}{(1-2\alpha,\alpha)}{(0,1)} .
\label{renewal:aPDF_early_FoxH}
\end{eqnarray}
\end{subequations}

From any of these representations, we learn that leading order corrections to the non-aged PDF are of the form $(t_a/t)^\alpha$.
An intuitive reasoning for this can be given as follows. A slightly late 
observer generally has to wait for the forward recurrence time before seeing the 
first event. The corrections thus have to account for waiting times which start 
earlier than the beginning of the observation $t_a$, but reach far into the 
observation time window $[t_a,t_a+t]$. Note that the number of (still few) 
waiting times drawn before time $t_a$ is measured by $n(t_a)\sim t_a^\alpha$, 
while the probability for any of them to be statistically relevant during an 
observation of length $t$ is proportional to 
$\int_t^\infty\psi(t)\,\mathrm{d}t\simeq t^{-\alpha}$. The expected number of 
statistically relevant waiting times starting before but reaching into the time 
window $[t_a,t_a+t]$ therefore scales like $t_a^\alpha \cdot t^{-\alpha}$, and 
so do leading order corrections.

\begin{figure*}
\includegraphics{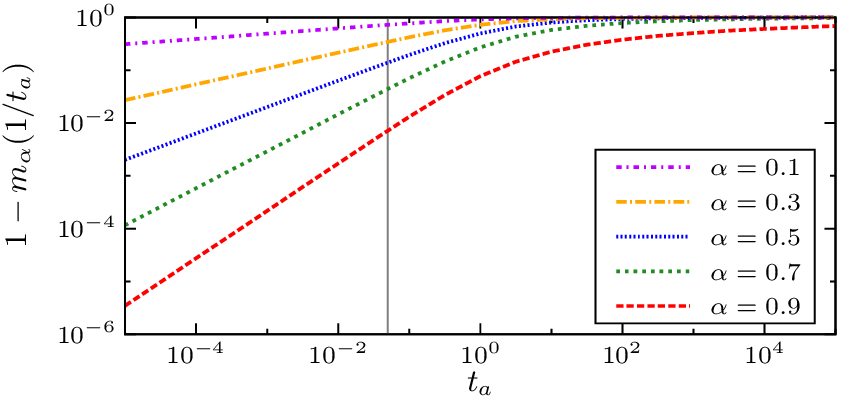}\includegraphics{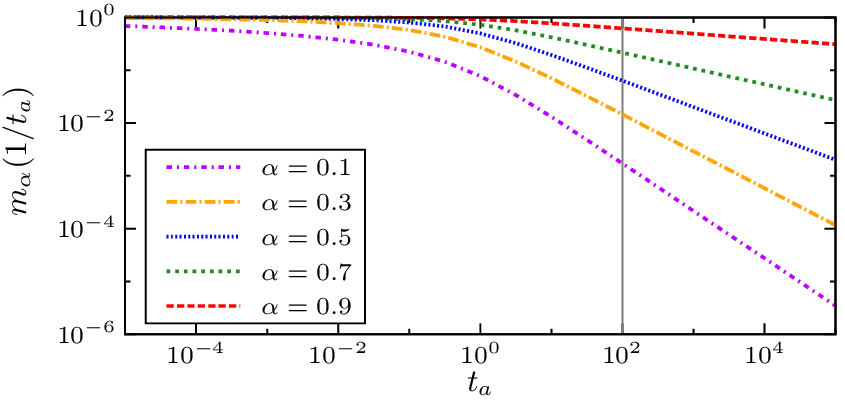}
 \caption{Double logarithmic plots of the probability $m_\alpha$ to observe any 
events during the measurement period $[t_a,t_a+t]$ (\textit{right}) and the 
complementary probability $1-m_\alpha$ (\textit{left}), both as a function of 
age $t_a$ at $t=1$. The full analytic behavior is given by 
Eqs.~(\ref{renewal:malpha}). Notice the initial power-law increase $(1-m_\alpha)
\simeq (t_a/1)^\alpha$ and the final power-law decay 
$m_\alpha\simeq (1/t_a)^{1-\alpha}$. Vertical lines indicate the values 
of $t_a$ used for the plots in Fig.~\ref{fig.PDF}.}
 \label{fig.malpha}
\end{figure*}

\begin{figure*}
\includegraphics{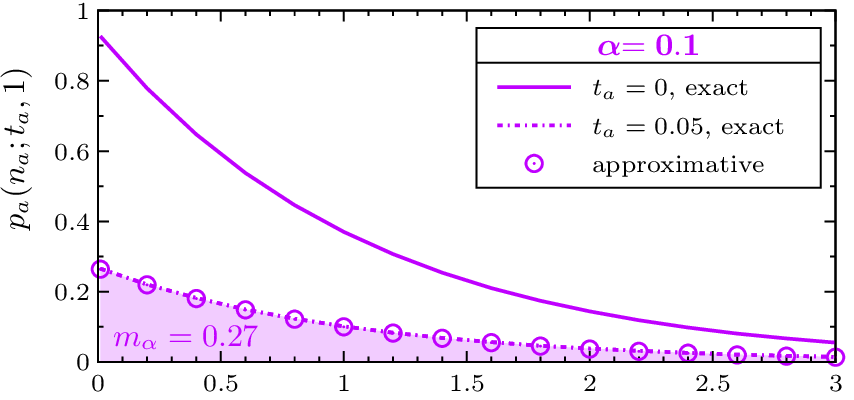}\includegraphics{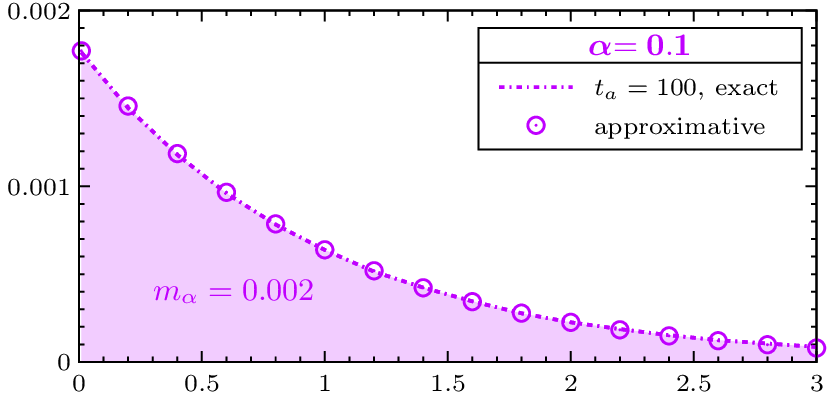}
\includegraphics{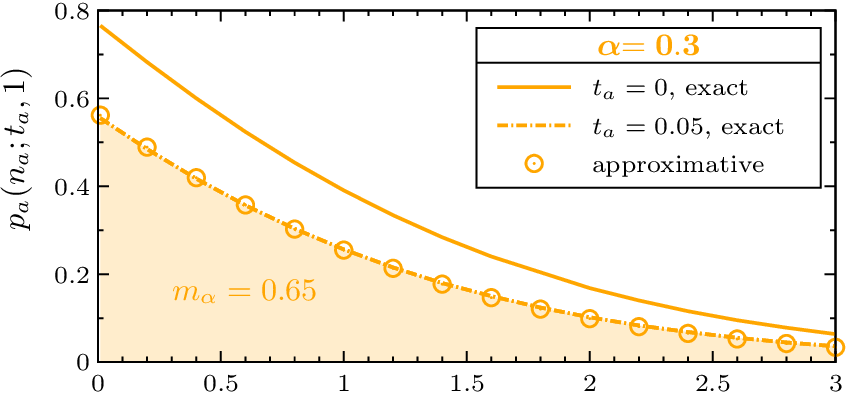}\includegraphics{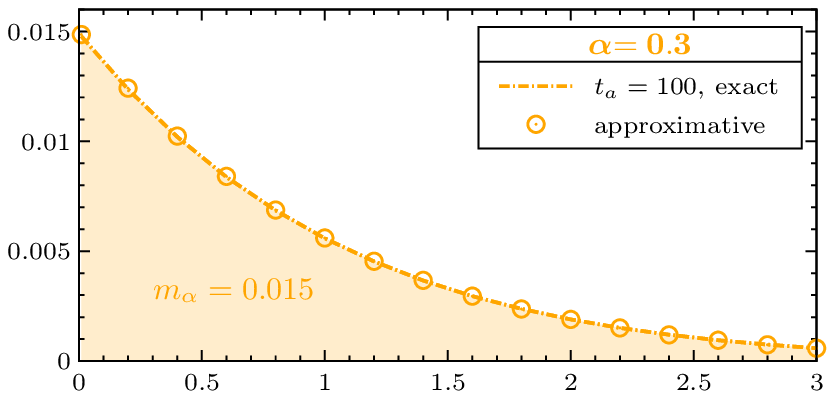}
\includegraphics{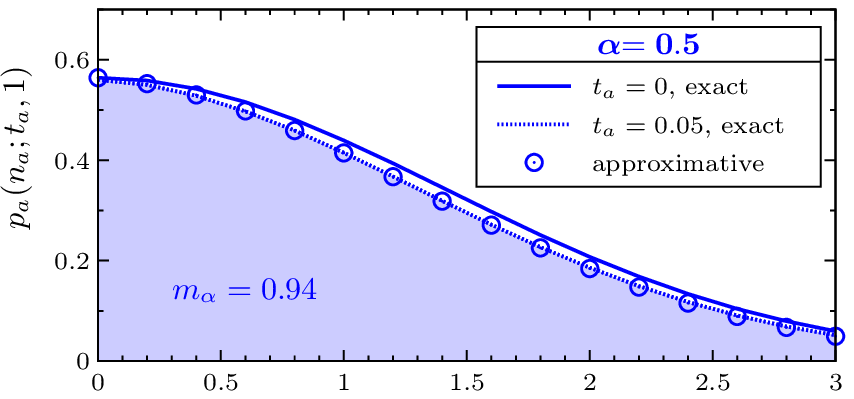}\includegraphics{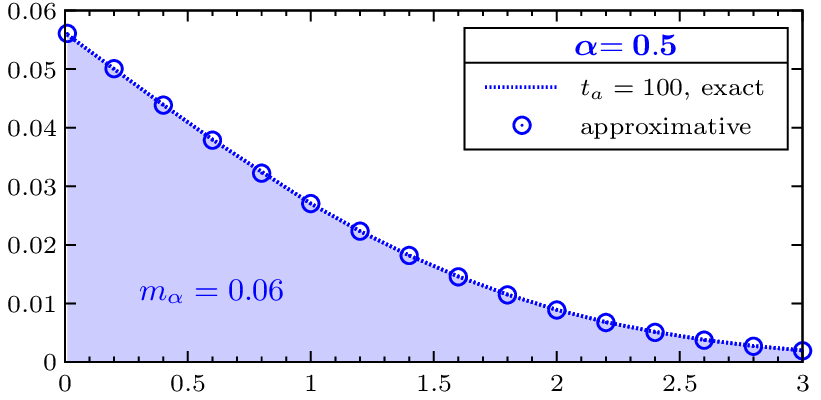}
\includegraphics{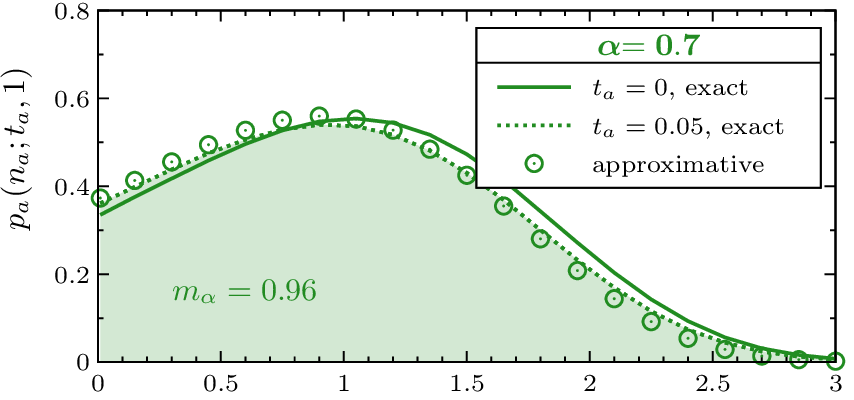}\includegraphics{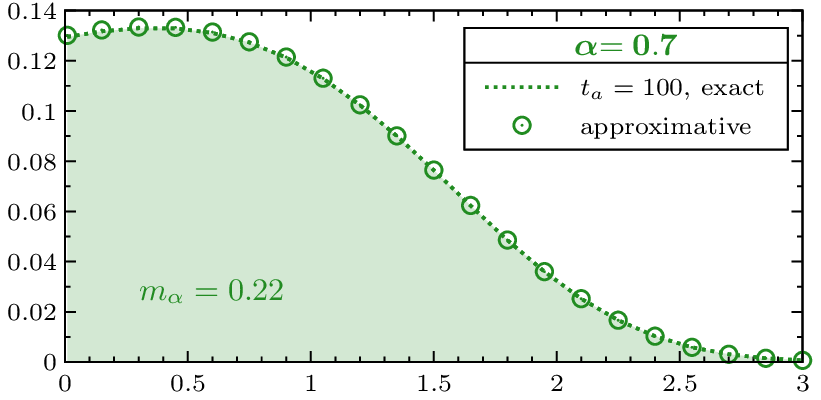}
\includegraphics{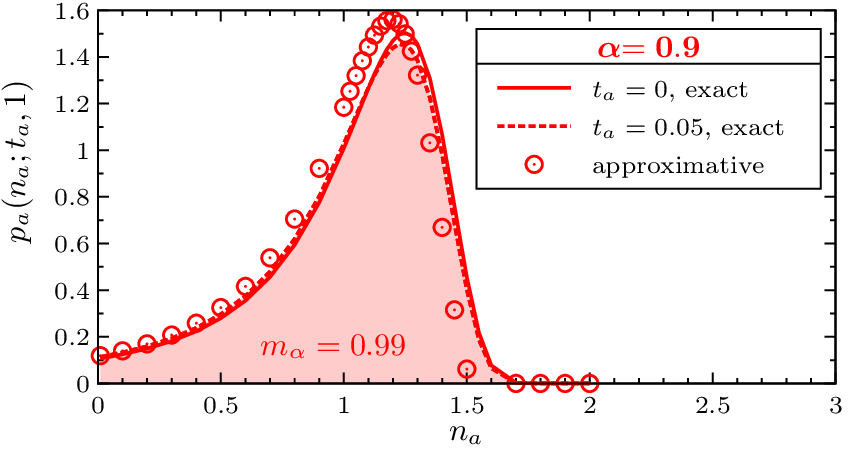}\includegraphics{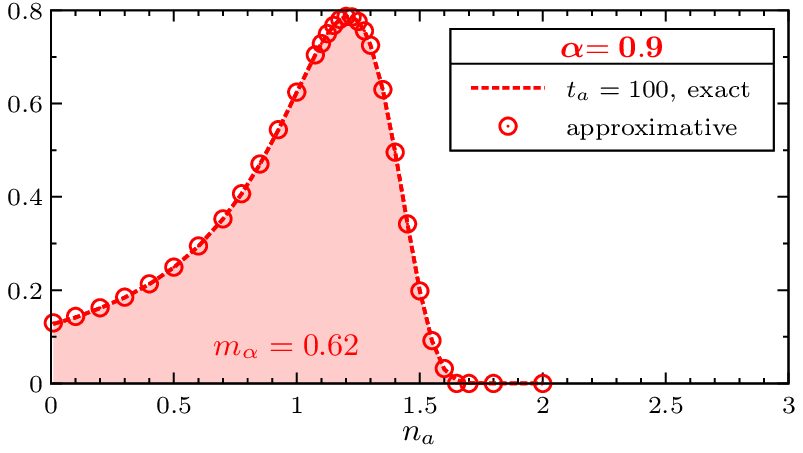}
\caption{Continuous part of the PDF  $p(n_a;t_a,t)$ of the number of events 
$n_a$ counted during the measurement period $[t_a,t_a+t]$. Exact results are provided in terms of a numerical evaluation of the convolution $h(t_a,t)*_t p(n_a;t)$ as defined through Eqs.~(\ref{renewal:pdf}) and (\ref{renewal:aPDFA}). We plot sample graphs for $t=1$ and various $\alpha$. Left panels show the non-aged case ($t_a=0$), the slightly aged case ($t_a=0.05$) and the leading order approximation for the slightly aged PDF [$t_a=0.05$, see Eq.~(\ref{renewal:aPDF_early_FoxH})]. Right panels show the highly
aged case ($t_a=100$) and the approximation thereto [$t_a=100$, see
Eq.~(\ref{renewal:aPDF_high_FoxH})]. Notice the significantly different vertical
scales in the panels, owing to the age-sensitive probability $m_\alpha$ 
to count any events at all during observation (cf.~Fig.~\ref{fig.malpha}). }
 \label{fig.PDF}
\end{figure*}

The precise nature of the modifications to the non-aged PDF due to aging can be 
separated into two aspects. On the one hand, the continuous part of the aged 
PDF, $h*_tp$, looses weight to the discrete $\delta(n_a)$-part with growing age 
$t_a$ of the counting process. This reflects an increasing probability to have 
waiting times that not only reach into, but actually span the full observation 
time window, so that no events at all are observed. We provide a graphical study 
of the early age-dependence of $1-m_\alpha$, i.e. the weight of the discrete contribution, in the left panel of Fig.~\ref{fig.malpha}. The double logarithmic plot clearly demonstrates the initial power-law growth $\simeq(t_a/t)^\alpha$. Moreover we see that for any fixed age $t_a/t$, the value of $1-m_\alpha$ is higher for lower values of $\alpha$. This was to be expected, since lower values of $\alpha$ are related to broader waiting time distributions, and thus stronger aging effects.

Interestingly, on the other hand, the modification to the continuous part of the aged PDF, $h*_tp$, goes 
beyond this weight transfer: It is not proportional to the non-aged PDF itself, 
but, according to Eq.~(\ref{renewal:aPDF_early_invstable}), to its slope. 
With increasing age $t_a/t$, regions with negative (positive) slope increase 
(decrease), so that local maxima have a tendency to shift towards $n_a=0$. 
Furthermore, we can deduce from the Fox-$H$ function representation, Eq.~(\ref{renewal:aPDF_early_FoxH}), the 
behavior around the origin, $0<n_at^{-\alpha}\ll1$, and the tail asymptotics, $n_at^{-\alpha}\gg 1$. See Appendix \ref{app:functions} for details. For $\alpha>1/2$, the 
initial slope of $h*_tp$ is negative, and hence the early aging effect is an increase of $h*_tp$ between the origin and the local maximum. This is notable since on the long run (i.e. for sufficiently long $t_a$), the probability to have any $n_a>0$ tends to zero. For $\alpha<1/2$, we find the converse: the initial slope is negative, and the PDF in the vicinity of the origin starts dropping from early ages. The PDF tails are, for any value of $\alpha$, of a compressed exponential form, meaning here $\log(h*_tp)\simeq n^{1/(1-\alpha)}$.

We can assess the validity of this early-age approximation by studying the left panels of Fig.~\ref{fig.PDF}. We plot the non-aged PDF, Eqs.~(\ref{renewal:pdf}) (\textit{full lines}) \cite{footnote:plotstable}, a numerical evaluation of the continuous part of the aged PDF, $h*_tp$, as given through the full convolution integral in Eq.~(\ref{renewal:aPDF}) (\textit{dotted/dashed lines}), and the early-age approximation by expanding the Fox-$H$ functions in Eq.~(\ref{renewal:aPDF_early_FoxH}) as power series (\emph{symbols}). All qualitative statements from the previous paragraph are confirmed by the sample plots. Still, the leading order approximation, Eqs.~(\ref{renewal:aPDF_early_LaplaceA}), is apparently not equally suitable for all values of $\alpha$. We observe deviations when $\alpha$ gets close to $1$. Indeed, one can show that the leading order terms in Eqs.~(\ref{renewal:aPDF_early_LaplaceA}), which  account for corrections of $\mathcal{O}([t_a/t]^\alpha)$, are followed by higher order terms of $\mathcal{O}(t_a/t)$. Hence, in general, these `almost leading order' corrections need to be taken into account when $\alpha$ is close to 1. Yet again, in this case, the total corrections with respect to the non-aged PDF, as measured in terms of the weight loss $1-m_\alpha$, are relatively small anyway.

\textit{Highly aged PDF. ---}
Conversely, we approximate for $s_a\ll s$: $h(s_a,s)\sim s_a^{-\alpha} 
s^{\alpha-1}$. This yields the leading order behavior for highly aged 
measurements, $t_a\gg t$:
\begin{equation}
1-m_\alpha(t/t_a)\sim1-\frac{(t/t_a)^{1-\alpha}}{\Gamma(\alpha)\Gamma(2-\alpha)}  \end{equation} 
and
\begin{subequations}
\begin{equation}
h(t_a,t)*_t p(n_a;t)\sim\frac{1}{\Gamma(\alpha)}\frac{1}{t_a^{1-\alpha}} 
\Lti{s^{2\alpha-2}e^{-n_as^\alpha}}{s}{t}.\\
\end{equation}
\label{renewal:aPDF_high_Laplace}%
Above Laplace inversion can be related to the non-aged PDF~(\ref{renewal:pdf}) through
\begin{align}
    h(t_a,t)*_t p(n_a;t) \sim -\frac{1}{\Gamma(\alpha)}\frac{1}{t_a^{1-\alpha}} \int_0^t \frac{\partial p(n_a;t')}{\partial n_a} \, \mathrm{d}t' ,\nonumber\\
  \label{renewal:aPDF_high_invstabl}
\end{align}
or expressed as Fox-$H$ function by
\begin{align}
  h(t_a,t)*_t p(n_a;t) \sim 
    \frac{(t/t_a)^{1-\alpha}}{\Gamma(\alpha)}\frac{1}{t^\alpha}\, 
\FH{1}{0}{1}{1}{\frac{n}{t^\alpha}}{(2-2\alpha,\alpha)}{(0,1)} .
  \label{renewal:aPDF_high_FoxH}
\end{align}
  \label{renewal:aPDF_high_FoxHA}
\end{subequations}

Again, modifications account for waiting times that start before but reach into 
the time window of observation. But in this late time regime, the initiation of 
the renewal process already lies far in the past, so not all waiting times 
before $t_a$ have a realistic chance to do that. Instead, we assume again that 
statistically relevant pre-measurement waiting times need to be of the order of 
$t$ (implying a probability $\simeq t^{-\alpha}$). To estimate their number, 
note that they must follow events which occur roughly within a time period 
$[t_a-t,t_a]$. However, at this (late) stage of the renewal process, the average 
event rate has dropped to $(dn/dt)_{t\approx t_a}\sim t_a^{\alpha-1}$. Thus the 
expected number of waiting times in the (comparatively short) time period 
preceding the measurement we conjecture scales like $\sim t_a^{\alpha-1}t$. This 
heuristic line of argument explains why for highly aged measurements, we have 
leading order contributions of the order $t_a^{\alpha-1}t\cdot 
t^{-\alpha}=(t/t_a)^{1-\alpha}$.

Graphical examples for this regime are given in the right-hand panels of 
Figs.~\ref{fig.malpha} and~\ref{fig.PDF}. Here, the continuous part $h*_tp$ of 
the aged PDF is not proportional to the slope of the non-aged PDF, but, according to Eq.~(\ref{renewal:aPDF_high_invstabl}), rather
its time integral over the duration of the measurement. Moreover, through its 
Fox-$H$ function representation~(\ref{renewal:aPDF_high_FoxH}), we learn that the initial slope of the late age 
PDF is positive for $\alpha>2/3$ and negative if $\alpha<2/3$. The far tail 
behavior however persists at late aging stages, as we still find  
$\log(h*_tp)\simeq n^{1/(1-\alpha)}$. Notice that in the case of high ages, in Fig.~\ref{fig.PDF} realized as $t_a/t=100$, the leading order terms~(\ref{renewal:aPDF_high_Laplace}) are satisfactorily approximating the exact convolution~(\ref{renewal:aPDF}) for all values of $\alpha$.

\subsection{Aging ensemble averages}\label{renewal:averages}
From the non-aged PDF in Eq.~(\ref{renewal:pdfL}) one can derive the expected time 
behavior of any function $f$ of the number of events $n$ counted since the 
initiation of the renewal process:
\begin{align}
 \langle f(n(t))\rangle &= \int_0^\infty f(n) \, p(n;t) \,dn \nonumber\\
 &= \Lti{s^{\alpha-1}\Lt{f(n)}{n}{s^\alpha}}{s}{t}.
 \label{renewal:meanf}
\end{align}
We give concrete examples below. First, we ask how such an ensemble average is 
altered if evaluated for the aged counting process. For this, we substitute $n$ 
by $n_a$ and $p$ by $p_a$ and insert 
Eqs.~(\ref{renewal:aPDFA}) to (\ref{renewal:aPDF_high_FoxHA}):
\begin{subequations}
\begin{multline}
 \langle f(n_a(t_a,t))\rangle = \int_0^\infty f(n_a) \, p_a(n_a;t_a,t) \,dn_a \\
  = f(0)[1-m_\alpha(t_a/t)] + h(t_a,t)*_t \langle f(n(t))\rangle,
\label{renewal:ameanf}
\end{multline}
with the limiting behavior
\begin{multline}
  h(t_a,t)*_t \langle f(n(t))\rangle \sim \\ \sim
  \begin{cases}
    \langle f(n(t))\rangle + t_a^\alpha \langle 
f'(n(t))\rangle/\Gamma(1-\alpha), & t_a\ll t\\[0.2cm]
    t_a^{\alpha-1} \int_0^t \langle f'(n(t'))\rangle \,\mathrm{d}t' / 
\Gamma(\alpha), & t_a\gg t
  \end{cases}.
  \label{renewal:ameanf_lim}
\end{multline}
  \label{renewal:ameanf_limA}
\end{subequations}
These equations relate an ensemble average taken for the observation window 
$[t_a,t_a+t]$ to the respective quantity measured in $[0,t]$. Interestingly, the 
modifications due to aging are rather related to the ensemble average of the 
derivative of the observable, $f'(n)=(\partial f/\partial n)(n)$.

As an example we consider $q$th order moments of the number of renewals, 
$f(n)=n^q$, $q>0$. We find
\begin{subequations}
\begin{equation}
\langle n^q(t)\rangle=\Lti{s^{\alpha-1}\frac{\Gamma(q+1)}{s^{\alpha q+\alpha}}}{
s}{t}=A_0t^{\alpha q}
\end{equation}
and
\begin{eqnarray}
\nonumber
\langle n_a^q(t_a,t)\rangle&=&\Gamma(q+1)(t+t_a)^{\alpha q}\\
\nonumber
&&\times\frac{\ibe{[1+t_a/t]^{-1}}{1+\alpha q-\alpha}{\alpha}}{\Gamma(\alpha)
\Gamma(1+\alpha q-\alpha)}\\[0.2cm]
&\sim&\left\{\begin{array}{ll}
\langle n^q(t)\rangle+A_1t_a^\alpha t^{\alpha q-\alpha}, & t_a\ll t\\[0.2cm]
B_1t_a^{\alpha-1}t^{1-\alpha+\alpha q}, & t_a\gg t
\end{array}\right..
\label{renewal:nq}
\end{eqnarray}
The incomplete Beta function comes into play again by substituting 
$u=t'/(t'+t_a)$ in the convolution integral in 
expression~(\ref{renewal:ameanf}). The time-independent coefficients are given 
by
\begin{align}
 A_0 &= \frac{\Gamma(q+1)}{\Gamma(\alpha q+1)}, \qquad A_1 = 
\frac{\Gamma(q+1)}{\Gamma(1-\alpha)\Gamma(\alpha q +1-\alpha)} , \nonumber\\
 B_1 &= \frac{\Gamma(q+1)}{\Gamma(\alpha)\Gamma(\alpha q+2-\alpha)} . 
 \label{renewal:nq_coeff}
\end{align}
\end{subequations}

For the non-aged system, moments evolve like $t^{\alpha q}$, reflecting the 
characteristic scaling property of the renewal process, $n(t)\sim t^\alpha$. But 
for aged systems, $t_a>0$, the scaling is broken, and moments behave in a more 
complex fashion with respect to time. Only at very high ages, $t_a\gg t$, we can 
approximate again by a single power-law. When comparing the growth of the 
counting processes for the two limiting regimes, Eq.~(\ref{renewal:nq}) is 
somewhat ambivalent. At high ages, the probability for observing no events at 
all tends to one. Consequently, a prefactor $t_a^{\alpha-1}$ lets all moments 
decay to zero as $t_a$ goes to $\infty$. However note that for a fixed, large 
but finite value of $t_a$, the $t$-dependence is $\simeq t^{1-\alpha+\alpha q}$ (in accordance with~\cite{Meroz2010}), 
so the power-law exponent is actually larger than for the non-aged moments. In 
summary, at higher ages $t_a$, the absolute number $n_a$ of counted events 
drops, but it increases faster with measurement time $t$. In particular, 
consider the average number of events during observation, $q=1$: Counting from 
the start of the process, we observe a sublinear behavior $\langle 
n\rangle\simeq t^\alpha$; but at fixed, high age of the process, the average 
rate of events is approximately constant, $\langle n_a\rangle\simeq t$, like in 
a nonaging, Poisson type of renewal process.

Another useful expression is the Laplace transform of 
Eq.~\eqref{renewal:pdfL} with respect to the number of renewals, 
$n\rightarrow\lambda$, and respectively Eq.~\eqref{renewal:aPDFL} with 
$n_a\rightarrow\lambda$, which is obtained through the same techniques.
Thus,
\begin{subequations}
\begin{equation}
\langle\exp[-\lambda n(t)]\rangle=\Lti{\frac{s^{\alpha-1}}{s^\alpha+\lambda}}{
s}{t}=\mle{\alpha}{-\lambda t^\alpha},
\label{renewal:nexp}
\end{equation}
and
\begin{multline}
  \langle\exp[-\lambda \, n_a(t_a,t)]\rangle \sim \\ \sim
  \begin{cases}
    \langle\exp[-\lambda \, n(t)]\rangle \, \left[1-\lambda 
t_a^\alpha/\Gamma(1-\alpha)\right]    ,&  t_a\ll t  \\
    1 - (t/t_a)^{1-\alpha} \gmle{\alpha}{2-\alpha}{-\lambda t^\alpha} / 
\Gamma(\alpha)  ,&  t_a\gg t
   \end{cases},
\label{renewal:nexp_lim}
\end{multline}
\end{subequations}
where $E_\alpha$ and $E_{\alpha,2-\alpha}$ are (generalized) Mittag-Leffler 
functions (see Appendix \ref{app:functions}). Interestingly, here the early 
first order corrections due to aging do not significantly alter the 
$t$-dependence. At low age, the Mittag-Leffler function interpolates between 
$1-\text{const}\cdot t^\alpha$ for $t\ll\lambda^{-1/\alpha}$ and 
$\text{const}\cdot t^{-\alpha}$ at $t\gg\lambda^{-1/\alpha}$. At high age, the 
transition is from $1-\text{const}\cdot t^{1-\alpha}$ to $1-\text{const}\cdot 
t^{1-2\alpha}$.

\subsection{Conditional ensemble averages}\label{renewal:conditionalaverages}

To conclude this section, we address the question of how counting statistics 
change when we selectively evaluate only realizations of the process where 
$n_a>0$. This means we discard the data when no events happen during the 
complete time of observation $[t_a,t_a+t]$. This is, on the one hand, a relevant 
question when it comes to the application of renewal theory: An observer who is 
unaware of the underlying counting mechanism, might misinterpret realizations 
with $n_a=0$ as a separate, dynamically different process, since the statistics 
are so distinct from the remaining continuum $n_a>0$. A process realization 
during which no events occur at all might even not be visible to the observer in the 
first place. We give an example in the next section. On the other hand, 
this study also helps us to distinguish two aspects of the aging PDF: We 
neglect the effect of having single waiting times that cover the full 
observation window, leading to a weight transfer from the continuous to the 
discrete part of the PDF. Instead we specifically only account for waiting times 
that start before but finish during the period of observation, in order to 
understand the modifications of the continuous part beyond its loss of weight. 
To this intent, we look at the conditional ensemble average
\begin{multline}
 \langle f(n_a(t_a,t))\rangle_m \equiv \int_0^\infty f(n_a) p_a(n_a|n_a>0;t_a,t) 
\,dn_a \\
 \begin{aligned}
   &= \frac{h(t_a,t)*_t \langle f(n(t))\rangle}{m_\alpha(t/t_a)} \\
   &\sim\begin{cases}
     \langle f(n(t))\rangle, &t_a=0\\[0.2cm]
     t^{\alpha-1} \int_0^t \langle f'(n(v))\rangle \,\mathrm{d}v 
/\Gamma(2-\alpha), & t_a/t\rightarrow\infty
   \end{cases}.
 \end{aligned}
\end{multline}
For $t_a=0$, as mentioned above, counting of events starts instantly, so the 
restriction to $n_a>0$ is redundant. But in the limit of late ages, a possibly 
nonzero forward recurrence time affects the measurement. We find that ensemble 
averages conditioned to $n_a>0$ have a well defined, nontrivial limiting time 
dependence, even when $t_a/t$ tends to infinity. This is in contrast to the full 
ensemble, where at infinite ages, $\langle f(n_a(t_a,t))\rangle\rightarrow 
f(0)$. As an example consider again the time evolution of $q$th order moments, 
restricted to $n_a>0$. We find
\begin{subequations}
\begin{equation}
 \langle n_a^q(t_a,t))\rangle \rightarrow 0, \quad\text{ as }\quad 
t_a/t\rightarrow\infty ,
\end{equation}
but
\begin{equation}
\langle n_a^q(t_a,t))\rangle_m \sim
\begin{cases}
A_0 \,t^{\alpha q}, &t_a=0\\
C_0 \, t^{\alpha q}, & t_a/t\rightarrow\infty
\end{cases},
\label{renewal:nq_conditional}
\end{equation}
where $A_0$ is given by Eq.~(\ref{renewal:nq_coeff}) and
\begin{equation}
C_0=\frac{\Gamma(q+1)\Gamma(2-\alpha)}{\Gamma(\alpha q+2-\alpha)}.
\end{equation}
\end{subequations}
Thus, as opposed to the unrestricted ensemble measurement, conditional moments 
scale like $\sim t^{\alpha q}$ in \emph{both} non-aged and extremely aged 
systems. Note however, that prefactors are different, and a behavior deviating 
from a simple power-law is still observable at intermediate ages.

\section{Aging continuous time random walks}\label{ctrw}
The theory of continuous time random walks (CTRWs) directly builds on the 
concept of the renewal theory. We are thus able to view many of the intriguing features of this random motion, such as anomalous diffusion, population splitting and weak ergodicity breaking, in the light of the abstract analytical renewal theory ideas developed above. To do so, we start with a definition of the CTRW model in section~\ref{ctrw:process}, extending the renewal process by a random spatial displacement component. We then review results from previous work in the field, discussing in particular the aspects of population splitting (section~\ref{ctrw:populationsplitting}), anomalous diffusion and weak ergodicity breaking (section~\ref{ctrw:msd}). Concurrently, we add elements from our own recent discussion~\cite{Schulz2013}, aiming at relating these phenomena to the aging properties of the underlying renewal process. Population splitting can be traced back to the partially discrete nature of the aging renewal PDF. Studying ergodicity ultimately leads us to a general in-depth study of aging ensemble and time averages (section~\ref{ctrw:ergodicity}). We work out the fundamental differences between these two types of averages under aging conditions (weak ergodicity breaking), but also find interesting parallels at late ages (equivalence in time scalings). Finally, we apply the methods and formulae developed in the course of our discussion to analyze a stochastic process which generalizes previous CTRW models by additional confinement, friction and memory components in section~(\ref{ctrw:flectrw}).

\subsection{From aging renewal theory to aging continuous time random walks}\label{ctrw:process}
Consider a random walk process (one dimensional, for the sake of simplicity) in which steps do not occur at a fixed deterministic rate, but are instead separated by random, real valued waiting times. The idea is to model the random motion in a complex environment where sticking, trapping or binding reactions are to be taken into account. The 
processes we study here are CTRWs with finite-variance jump lengths on the one 
hand, and independent, identically distributed waiting times on the other (see 
\cite{report,Hughes} for a review on CTRW theory). Eventually we are interested in 
studying these processes on long time scales, aiming at extending existing 
simple diffusion models to describe diffusion in complex environments. Thus, we 
focus again on scale-free waiting time distributions of the 
form~(\ref{intro:tail}), which have the capability of modifying the resulting 
diffusion dynamics on arbitrarily long time scales.

In the simplest scenario, sticking or trapping mechanisms are decoupled from 
diffusion dynamics. On the level of theoretical modeling, this means we can 
take jump distances of the base random walk to be independent from waiting 
times. The renewal theory of the previous section is readily extended to 
describe such idealized systems. Let $x(n)$ be the random walk process as a 
function of the number of steps $n$. Then by means of subordination~\cite{Meerschaert2004,Fogedby1994,Baule2005,Magdziarz2007}, we 
construct a CTRW as $x(t)=x(n(t))$, where $n(t)$ is a renewal counting process. 
Each step of the random walk is hence interpreted as an event of the renewal 
process. The properties of $x(t)$ follow from the combined statistics of $x(n)$ 
and $n(t)$. For example, if $W_{\text{RW}}(x;n)$ denotes the PDF for the 
position coordinate $x$ after $n$ steps, starting with $x(0)=0$, then 
\begin{equation}
  W_{\text{CTRW}}(x;t) = \int_0 ^\infty W_{\text{RW}}(x;n) \, p(n;t) \,dn
 \label{ctrw:subordination}
\end{equation}
is the PDF of the associate CTRW process at time $t$. Note that we see $n$ as a 
continuous variable here, so we are arguing on the level of long time scaling 
limits. In the simplest case, $x(n)$ would be an ordinary Brownian motion so 
that $W_{\text{RW}}$ would be a Gaussian.

Now imagine that a particle is injected into a complex environment, beginning a CTRW 
like motion at time $0$. In general, the experimentalist might start its 
observation at a later time $t_a>0$. The reason for this could be experimental 
limitations, or maybe the goal is to study a maximally relaxed system, which makes 
it necessary to wait for relaxation. In either case, the particle motion is 
initially unattended. At time $t_a$ the particle is tracked down, and its 
position at this instant serves as the origin of motion for the following 
observations. Instead of the full CTRW $x(t)$, the experimentalist then monitors 
the aged CTRW $x_a(t_a,t)=x(t_a+t)-x(t_a)$. Thus, as we worked out in the 
previous section, if the dynamics are characterized by heavy-tailed trapping or 
sticking times, the issue of aging has to be taken into careful consideration.

\subsection{Population splitting}\label{ctrw:populationsplitting}
Arguably the most striking aging effect in CTRW theory is the emergence of an 
apparent population splitting~\cite{Schulz2013,Barkai2003b}. The aged renewal process  $n_a(t_a,t)$ controls the dynamic activity of the aged CTRW $x_a(t_a,t)$.  Consequently, the forward recurrence time $t_1$ marks the onset of dynamic  motion in the monitored window of time. We learned from the analysis of aging  renewal theory that for increasingly late measurements, we should expect $t_1$ to assume considerable values. Physically, this reflects the possibility for the particle to find ever deeper traps or to undergo more complex binding procedures, 
when given more time to probe its environment before the beginning of  observation. In particular, the forward recurrence time is more and more likely 
to span the full observation time window, $t_1>t$, so that the particle does not 
visibly exhibit dynamic activity.

From an external, experimental point of view, an ensemble measurement in such a
system appears to indicate a splitting of populations. Let, for instance, the 
base random walk $x(n)$ be Markovian, and let the PDF $W_{\text{RW}}(x;n)$ be a 
continuous function of $x$ (e.g. Gaussian). In this case we can supplement the 
ensemble PDF $W_{\text{CTRW}}$ from Eq.~(\ref{ctrw:subordination}) by its aged 
counterpart $W_{\text{ACTRW}}$, using renewal theory Eq.~(\ref{renewal:ameanf}):
\begin{eqnarray}
\nonumber
&&W_{\text{ACTRW}}(x_a;t_a,t)\\[0.2cm]
\nonumber
&&\hspace*{0.8cm}=\int_0 ^\infty W_{\text{RW}}(x_a;n_a)p_a(n_a;t_a,t)dn_a\\[0.2cm]
\nonumber
&&\hspace*{0.8cm}=\delta(x_a) [1-m_\alpha(t_a/t)]\\[0.2cm]
&&\hspace*{1.8cm} + h(t_a,t)*_t W_{\text{CTRW}}(x_a;t).
 \label{ctrw:subordination_aged}
\end{eqnarray}
This propagator was discussed previously in~\cite{Barkai2003b}. The ensemble statistics have a sharply peaked $\delta$-contribution, caused by a 
fraction $1-m_\alpha$ of particles which remain immobile at the origin of the 
observed motion. They contrast the mobile fraction $m_\alpha$ of particles, 
since the PDF of their arrival coordinates, $h*_tW_{\text{CTRW}}$, varies 
continuously along the accessible regions of space $0<x_a<\infty$ (as derived, 
by virtue of Eq.~(\ref{ctrw:subordination}), from the continuous nature of the 
PDFs $h$, $W_{\text{RW}}$ and $p$). For fixed evaluation time $t$, the size of 
the immobile subpopulation increases with growing age $t_a$, as aging renewal 
dynamics terminally come to a complete halt. An exhaustive discussion on the 
shape of the aged propagator $W_{\text{ACTRW}}$ can be found in~\cite{Barkai2003b} for both unbiased motion and in the presence of a drift.

Indeed, splitting into subpopulations is a phenomenon encountered in complex environments such as biological cells. Such dynamics was observed for the motion of lipids in phospholipid membranes~\cite{Schutz1997}, single protein molecules in the cell nucleus~\cite{Kues2001}, H-Ras on plasma membranes~\cite{Lommerse2005}, and of membrane proteins~\cite{Manley2008}. The immobile fraction is also often found in fluorescence recovery after photobleaching experiments~\cite{Schutz1997}.

It is important to understand, that for CTRW types of motion, this effect emerges without assuming  non-identical particle dynamics. Even during the evolution of the process, stochastic motion of individual particles in an ensemble is independent and identical. In particular, all particles in principle exhibit their dynamic activity for an indefinite amount of time. The `immobility attribute' can only be assigned when the evolution of the (aged) process is studied within a \emph{finite time} window. Then, a certain fraction of particles---the immobile ones---stand out \emph{statistically} from the rest. The displacement propagator~(\ref{ctrw:subordination_aged}) with its conspicuous, discontinuous contribution serves as a statistical indicator for the population splitting, if evaluated at finite times $t<\infty$ (more precisely, the effect is most noticeable while $t$ remains short as compared to the age $t_a$). Similarly, we show in the following section that population splitting is particularly  relevant when assessing time average data on a per-trajectory basis. Of course, 
in any case, observations of real physical systems are finite by nature. It is 
hence important to know the characteristics of the aging population splitting, 
as to set it apart from separation mechanisms due to physically non-identical 
particle dynamics.

\subsection{Analysis of mean squared displacements}\label{ctrw:msd}

An alternative way to assess particle spreading in the solvent is to study the 
time evolution of the mean squared displacement. This is particularly useful 
when ensemble data is not extensive enough as to deduce reliable propagator 
statistics $W_{\text{ACTRW}}$. There are two common ways of defining such mean 
squared displacements: either in terms of an ensemble average or a 
single-trajectory time average. Analysis and comparison of these two types of 
observables reveals several fingerprint phenomena of CTRW motion, such as 
subdiffusion and weak ergodicity breaking. In the following, we collect 
known results and discuss their implications for aged systems.

\textit{Ensemble average. ---}
As a simple example, we take $x(n)$ to be unbounded, unbiased Brownian motion 
and consider an ensemble measurement of the mean squared displacement. In this 
case, the PDF $W_\text{RW}$ is Gaussian, and we know that
\begin{eqnarray}
\nonumber
\langle[x(k+n)-x(k)]^2\rangle&=&\langle x^2(n)\rangle\\
\nonumber
&=&\int_{-\infty}^{\infty}x^2W_\text{RW}(x;n)\mathrm{d}x\\
&=&2Dn
\label{ctrw:eamsd_bm}
\end{eqnarray}
for all $k,n>0$. We use arbitrary spatial units from here, $2D=1$. Since $x(n)$ 
is a process with stationary increments, the calculation of moments is 
independent of the number of steps $k$ made before start of the measurement. 
Thus, if the process the experimentalist studies is Brownian motion, there are no aging effects: at all initial times $k$, one observes a linear scaling 
with respect to observational time $n$, $\langle x^2(n)\rangle\simeq n$, a 
behavior commonly classified as normal diffusion. However, if the motion is 
paused irregularly for heavy-tailed waiting time periods, the dynamics are 
of CTRW type and we get a quite different picture. For the process 
$x(t)=x(n(t))$, we find
\begin{eqnarray}
\nonumber
\langle [x(t_a+t)-x(t_a)]^2\rangle&=&\langle x_a^2(t_a,t)\rangle\\
\nonumber
&&\hspace*{-3.8cm}
=\int_{-\infty}^{\infty}x_a^2W_\text{ACTRW}(x_a;t_a,t)\mathrm{d}x_a\\
\nonumber
&&\hspace*{-3.8cm}
=\int_{-\infty}^{\infty}\int_0^\infty x_a^2W_\text{RW}(x_a;n_a)p_a(n_a;t_a,t)
\mathrm{d}n_a\mathrm{d}x_a\\
\nonumber
&&\hspace*{-3.8cm}
=\int_0^\infty n_ap_a(n_a;t_a,t)\mathrm{d}n_a\\ 
\nonumber
&&\hspace*{-3.8cm}
=\langle n_a(t_a,t)\rangle=\langle n(t_a+t)\rangle-\langle n(t_a)\rangle\\
\nonumber
&&\hspace*{-3.8cm}
=\frac{1}{\Gamma(1+\alpha)}\left[(t_a+t)^\alpha-t_a^\alpha\right]\\
&&\hspace*{-3.8cm}
\sim\left\{\begin{array}{ll}
t^\alpha/\Gamma(\alpha+1)+t_a^\alpha/\Gamma(1-\alpha), &t_a\ll t\\[0.2cm]
t_a^{\alpha-1}t/\Gamma(\alpha), &t_a\gg t\end{array}\right.,
\label{ctrw:eamsd}
\end{eqnarray}
in accordance with~\cite{Barkai2003b}.
As expected, the nonstationarity of the aging renewal process carries over to 
the CTRW. If monitored at $t_a=0$, the mean squared displacement grows like 
$\langle x_a^2\rangle=\langle x^2\rangle\simeq t^\alpha$. With the increase 
being less than linear in time, the phenomenon in the context of diffusion 
dynamics is commonly referred to as subdiffusion. For $t_a>0$, the mean squared 
displacement is no longer described in terms of a single power-law. The process 
looks more like diffusion in a nonequilibrium environment and in fact $t_a$ 
might be conceived as an internal relaxation time scale. Interestingly, if the 
measurement takes place in the highly aged regime $t_a\gg t$, the mean squared 
displacement as a function of the observation time $t$ indicates normal 
diffusion with an age dependent diffusion coefficient, $\langle 
x_a^2\rangle\simeq t_a^{\alpha-1}t$. Only if data on an extensive range of time 
scales is available, we are able to identify the complete turnover from one 
aging regime to the other as a fingerprint of CTRW dynamics. Still, for as long 
as the experimentalist cannot control the age $t_a$ of the measurement, the 
aging effect can be misinterpreted as an internal relaxation mechanism. The 
situation gets even more complicated, when $t_a$ is possibly random.

\textit{Time average. ---}
Now consider the alternative time average notion of the mean squared 
displacement. For a single particle trajectory $x(t)$, recorded at times 
$t'\in[t_a,t_a+T]$, it is defined in terms of the sliding average
\begin{equation}
\overline{\delta^2(\Delta;t_a,T)}=\frac{1}{T-\Delta}\int_{t_a}^{t_a + 
T-\Delta}{\left[x(t'+\Delta)-x(t')\right]^2}\mathrm{d}t'.
\label{ctrw:tamsd}
\end{equation}
Here, $\Delta$ is called lag time, and parameters defining the time window of 
observation are also referred to as age $t_a$ and measurement time $T$. While the 
ensemble mean~(\ref{ctrw:eamsd}) is evaluated in terms of squared displacements 
from a multitude of independent process realizations, the time 
average~(\ref{ctrw:tamsd}) uses data from within a single trajectory at several 
points in time. The latter is thus a useful alternative whenever measurements on 
long (i.e. $T\gg\Delta$) but relatively few trajectories are available. For 
ergodic, stationary processes, both types of averages are equivalent. For 
example, for a Brownian motion, the time average $\overline{\delta^2}$ is by 
definition a random quantity differing from one trajectory to the next; but in 
the limit of long trajectory measurements, $T\gg\Delta$, the time average 
converges to the corresponding ensemble value, $\overline{\delta^2}\rightarrow 
2D\Delta$, and fluctuations become negligible \cite{Burov2010}.

\begin{figure}
 \includegraphics{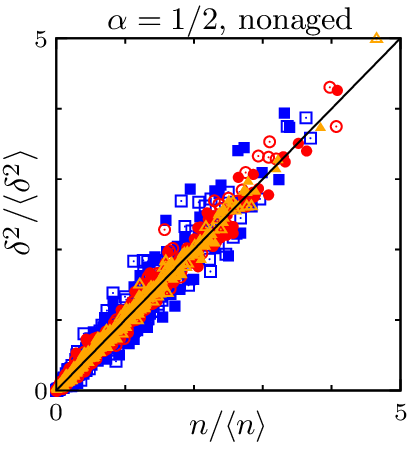} \includegraphics{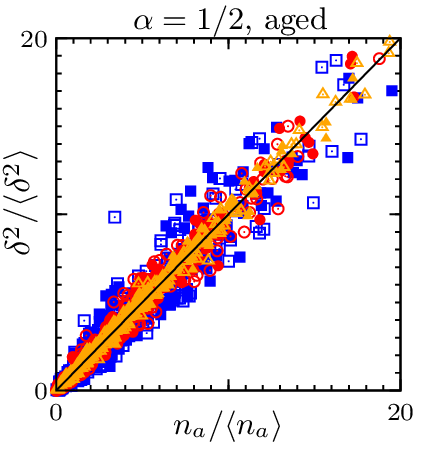}
 \includegraphics{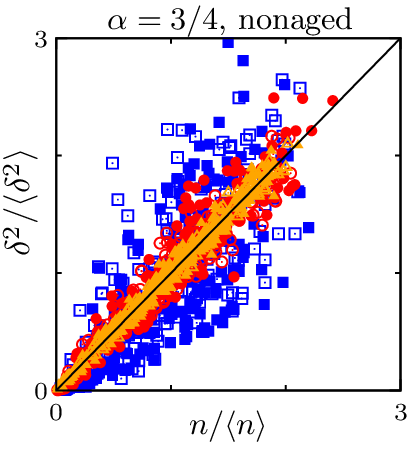} \includegraphics{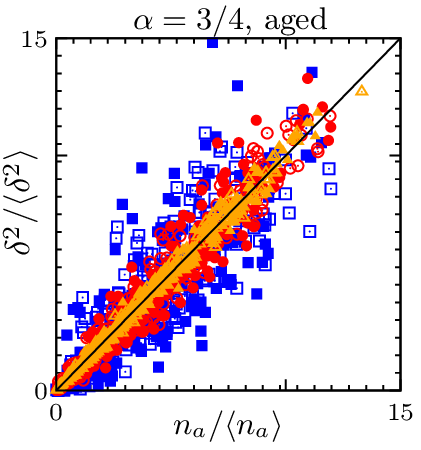}
 \caption{
 Numerical demonstration of the asymptotic identity~(\ref{ctrw:tamsd_pdf}) in the limit of long measurement times $T\gg\Delta$. While $\Delta=100$ remains fixed, we compare $T=2\times10^5$ (blue squares~\textcolor{blue}{$\blacksquare$}), $T=2\times10^6$ (red discs~\textcolor{red}{$\bullet$}), and $T=2\times10^7$ (orange triangles~ \textcolor{gnu4}{$\blacktriangle$}), showing nice convergence. Each point in the graph represents an individual CTRW trajectory. Full symbols represent free CTRW, open symbols are for motion bounded by a box. $t_a$ is either $0$ (non-aged), or for specific $\alpha$ and $T$ chosen such that $m_\alpha(t_a/T)=0.21$ (aged). The simulation data is extensive enough as to ensure that every system is represented by roughly $200$ points with $\overline{\delta^2}>0$. }
 \label{fig.n_vs_delta2}
\end{figure}

In contrast, the CTRW $x(t)=x(n(t))$ violates both ergodicity 
($\overline{\delta^2}$ remains random for arbitrarily long measurement times) 
and stationarity ($\overline{\delta^2}$ depends on $t_a$ and $T$). In the 
context of aging, we can now ask two questions. First, does the distribution of 
the time average change qualitatively when evaluated after the onset of the particle 
dynamics, $t_a>0$? This issue is discussed extensively in~\cite{Schulz2013}. In short, 
we find that the time averaged mean squared displacement is directly related to 
the number of steps $n_a$ made during the measurement via~\cite{He2008,Burov2010,Jeon2010}
\begin{equation}
\frac{\overline{\delta^2(\Delta;t_a,T)}}{\left<\overline{\delta^2(\Delta;
t_a,T)}\right>}=\frac{n_a(t_a,T)}{\langle n_a(t_a,T)\rangle} .
\label{ctrw:tamsd_pdf}
\end{equation}
Figure~\ref{fig.n_vs_delta2} provides a numerical validation of this relation in terms of explicit CTRW simulations for free and confined motion. Notice that Eq.~(\ref{ctrw:tamsd_pdf}) is not a distributional equality, but meant to be a stronger, per-trajectory equality which we validate here by means of simulations data. Hence, all random properties of the time average in this case are not only related to, but quite literally taken over from the underlying counting process. In particular, the distribution of the time average 
$\overline{\delta^2}$ is a rescaled version of the renewal theory PDF $p_a$ as 
discussed in section~\ref{renewal} and plotted in Figs.~\ref{fig.malpha} 
and~\ref{fig.PDF}. This implies that a statistical study of time 
averages reveals the aging population splitting: within an ensemble of 
independent particles, we may find some which do not exhibit any dynamic 
activity during observation, $n_a=0$ or $\overline{\delta^2}=0$, and thus apparently 
stand out as an individual subpopulation from the remaining continuum, 
$n_a>0$ or $\overline{\delta^2}>0$. 
Likewise, one can study~\cite{Albers2013} the statistics of microscopic diffusivities $D_\alpha=\delta^2/\Delta^\alpha=[x(t'+\Delta)-x(t')]^2/\Delta^\alpha$ along a single time series $t'\in[0,T]$. These diffusivities are consistently found to have a discrete probability for $D_\alpha=0$. The latter implies that no steps are made during any $\Delta$-sized lag time interval along the time series, and the probability for this actually grows with measurement time $T$.

The second aging-related question addresses the explicit lag time dependence. 
Albeit being different from one trajectory to the next, $\overline{\delta^2}$ is 
generally characterized by a universal scaling with respect to lag time $\Delta$. 
The question now is: Is this scaling age-dependent? To this end, we consider
the average (over many individual trajectories) of Eq.~(\ref{ctrw:tamsd}). By
help of Eq.~(\ref{ctrw:eamsd}),
\begin{subequations}
\begin{multline}
\left<\overline{\delta^2(\Delta;t_a,T)}\right>= \\
 \begin{aligned}
 &= \frac{1}{T- \Delta} \int_{t_a} ^{t_a +T-\Delta}{\langle\left[ x(t'+ 
\Delta)-x(t') \right]^2\rangle  } \,\mathrm{d}t' \\
 &=\frac{1}{(T- \Delta)\,\Gamma(2+\alpha)} \left[ (t_a+T)^{\alpha+1} \right. \\
 &\quad\quad \left. -(t_a+T-\Delta)^{\alpha+1}-(t_a+\Delta)^{\alpha+1} 
+t_a^{\alpha+1} \right] \\
 &\sim \frac{\Lambda_\alpha(t_a / T)}{\Gamma(1+\alpha)} \; 
\frac{\Delta}{T^{1-\alpha}} ,
 \label{ctrw:eatamsd}
 \end{aligned}\\
\end{multline}
where
\begin{equation}
\Lambda_\alpha(z) = (1+z)^\alpha - z^\alpha.
\label{ctrw:depression}
\end{equation} 
\end{subequations}
The approximation in the fourth line is for the relevant case of long 
measurement times, $T\gg\Delta$. A few points are remarkable when comparing the 
time dependence of the ensemble average, Eq.~(\ref{ctrw:eamsd}) to the scaling 
of the time average, Eq.~(\ref{ctrw:eatamsd}). The answer to the question on the
lag time dependence is a simple one: We generally find a linear scaling 
$\overline{\delta^2}\sim\Delta$, regardless of age $t_a$. In this respect, the 
time average is a less complicated observable than the ensemble average. The 
latter has a comparatively complicated $t$-dependence and is characterized by an 
age-dependent regime separation. Compare this to the time average, where 
nonstationarity enters in terms of the prefactors $\Lambda_\alpha$ and 
$T^{\alpha-1}$. Only the amplitude of time averages depends on the measurement 
duration and age. More precisely, if either measurement time parameter tends to 
$\infty$, the time average itself tends to zero. This is why we call 
$\Lambda_\alpha$ an \textit{aging depression}. It can be expressed in terms of 
the dimensionless ratio $t_a/T$. In analogy to ensemble measurements we thus speak of a non-aged ($t_a=0$), a slightly aged ($t_a\ll T$) and a highly 
aged ($t_a\gg T$) time average.

Aside from the differences, there is also an interesting parallel between 
ensemble and time averages: the linear scaling of the time average with respect 
to $\Delta$ is reminiscent of the linear $t$-dependence of the ensemble average 
at high ages. If, in addition to a very long measurement duration, we also 
assume an even longer preceding aging period, $t_a\gg T\gg \Delta$, the 
similarity even becomes an equivalence,
\begin{equation}
\left<\overline{\delta^2(\Delta;t_a,T)}\right>\sim t_a^{\alpha-1}\frac{\Delta}{
\Gamma{(\alpha)}}\sim\langle x_a^2(t_a,\Delta)\rangle .
\end{equation}
The time scalings of ensemble and time averages are hence identical at high ages. This is quite surprising considering that these quantities are fundamentally distinct in a process that exhibits weak ergodicity breaking. In what follows, we  
discuss whether or not these discrepancies and parallels of the two types of 
averages are specific to the mean squared displacement.

\subsection{Aging ensemble and time averages}\label{ctrw:ergodicity}
Consider a random walk $x(n)$ and a stationary observable $F(x_2,x_1)$, meaning 
that
\begin{equation}
 \langle F(x(n+k),x(k)) \rangle = \langle F(x(n),x(0)) \rangle \equiv f(n)
 \label{ctrw:stationarity}
\end{equation}
for any number of steps $n$ or $k$. In other words, the above ensemble average 
should not depend on when we begin the observation. The example discussed in the 
previous section falls into this category. There, $x(n)$ was Brownian motion, 
$F(x_2,x_1)=(x_2-x_1)^2$ was the squared displacement, and we had $f(n)=2Dn$. 
But condition~(\ref{ctrw:stationarity}) would also allow for general order 
moments or any other function of displacements $F(x_2-x_1)$. Moreover, we can 
replace $x(n)$ by other processes with stationary increments, like fractional 
Brownian motion~\cite{Mandelbrot1968}.  If $x(n)$ is even stationary itself (e.g. the stationary limit of confined motion), then any function $F$ is fine (allowing to calculate, e.g. correlation functions, $F(x_2,x_1)=x_2x_1$; CTRW multi-point correlation functions have been studied extensively in~\cite{Baule2005,Sanda2005,Barkai2007,Burov2010}).

\textit{Ensemble averages. ---}
In any such case, we can ask the question of how the statistical properties of 
the random motion and the measured observable change when introducing 
heavy-tailed waiting times between steps. We define $x(t)=x(n(t))$ via 
subordination, assuming that $x(n)$ and $n(t)$ are stochastically independent 
processes. In general, the aging properties of the renewal process $n(t)$ are inherited by the CTRW $x(t)$. The stationarity of the ensemble average is broken.
To calculate the magnitude of the effect, we can use conditional 
averaging by virtue of the independency of the two stochastic processes at work. We denote by $\langle\cdot\rangle_\mathrm{RW}$ the average with respect to realizations $x(n)$ and write
\begin{subequations}
\begin{eqnarray}
 &&\langle F(x(t_a+t),x(t_a))\rangle = \langle F(x(n(t_a+t)),x(n(t_a)))\rangle 
\nonumber\\
 &&\quad= \int_0^\infty\int_0^\infty \langle F(x(n_2),x(n_1))\rangle_{\mathrm{RW}} \nonumber\\
 &&\qquad\times\mathrm{Pr}\{n(t_a)=n_1,n_a(t_a,t)=n_2-n_1\} \,\mathrm{d}n_1\mathrm{d}n_2 \nonumber\\
 &&\quad= \int_0^\infty\int_0^\infty f(n_2-n_1)\nonumber\\
 &&\qquad\times\mathrm{Pr}\{n(t_a)=n_1,n_a(t_a,t)=n_2-n_1\} \,\mathrm{d}n_1\mathrm{d}n_2 \nonumber\\
 &&\quad= \int_0^\infty\int_0^\infty f(m)\mathrm{Pr}\{n(t_a)=n_1,n_a(t_a,t)=m\} \,\mathrm{d}n_1\mathrm{d}m \nonumber\\
 &&\quad= \int_0^\infty f(m) p_a(m;t_a,t)\mathrm{d}m \nonumber\\
 &&\quad= \langle f(n_a(t_a,t))\rangle .
 \label{ctrw:ea}
\end{eqnarray}
The average on the last line is with respect to $n_a(t_a,t)$, so we can use our knowledge on ensemble averages of the aging renewal 
process. The latter are characterized by a distinct turnover behavior, see 
Eqs.~(\ref{renewal:meanf}) and (\ref{renewal:ameanf_limA}). For the slightly aged 
CTRW ensemble average, we have
\begin{align}
 \langle f(n(t))\rangle &=  \Lti{s^{\alpha-1}\Lt{f(n)}{n}{s^\alpha}}{s}{t} , 
\nonumber\\
 \langle f(n_a(t_a,t))\rangle &= \langle f(n(t))\rangle + 
\mathcal{O}((t_a/t)^\alpha) \quad (t_a\ll t) .
 \label{ctrw:ea_low}
\end{align}
For CTRWs, the leading order corrections due to aging are of the order 
$(t_a/t)^{\alpha}$, just as for the underlying renewal process. Conversely, at 
high ages, we slightly rewrite the leading order terms as
\begin{multline}
 \langle f(n_a(t_a,t))\rangle \sim \\
 \begin{aligned}
   &\sim f(0) + \frac{t_a^{\alpha-1}}{\Gamma(\alpha)} \int_0^t \left[ \langle 
f'(n(t'))\rangle -f(0)\,\frac{(t')^{-\alpha}}{\Gamma(1-\alpha)} \right] 
\,\mathrm{d}t' \\
   &= f(0) + \frac{t_a^{\alpha-1}}{\Gamma(\alpha)} \; 
\Lti{s^{2\alpha-2}\Lt{f(n)-f(0)}{n}{s^\alpha}}{s}{t} \\
   &\equiv C + \frac{t_a^{\alpha-1}}{\Gamma(\alpha)} \; g(t) \qquad\qquad\qquad 
(t_a\gg t),
 \end{aligned}
\label{ctrw:ea_high}
\end{multline}
\label{ctrw:ea_highA}
\end{subequations}
introducing the constant $C=f(0)$ and defining an auxiliary function $g(t)$, 
either relating it to the analog stationary average $f(n)$ (third line) or to 
the corresponding non-aged CTRW average $\langle f'(n(t))\rangle$ (second line). 
When the measurement of the observable $F$ is taken arbitrarily late, 
$t_a\rightarrow\infty$, we will ultimately measure the constant $C$. For 
example, if the base random motion $x(n)$ is a random walk with a characteristic 
scaling $x\sim n^H$, $H>0$, and we study moments of displacements, 
$F(x_2,x_1)=|x_2-x_1|^q$, $q>0$, then $f(n)\simeq n^{qH}$; in this case, $C=0$, 
so we will get arbitrarily small moments at high ages. If $x(n)$ is the 
stationary limit of a confined motion and we are interested in the correlation 
function $F(x_2,x_1)=x_2x_1$ (as studied in~\cite{Burov2010,Schulz2013}), then late measurements will be close to the 
thermal value of the squared position $C=\langle x^2\rangle$. The decay to this 
constant value is generally of the order $(t/t_a)^{1-\alpha}$, no matter which 
observable we are studying.

\textit{Time averages. ---}
Now we turn to the analog time average, namely
\begin{subequations}
\begin{equation}
\overline{F(\Delta;t_a,T)}= \frac{1}{ T- \Delta} \int_{t_a} ^{t_a + T-\Delta} 
F(x(t'+\Delta),x(t')) \,\mathrm{d}t' ,
\end{equation}
and calculate its expectation value
\begin{eqnarray}
\nonumber
\left<\overline{F(\Delta;t_a,T)}\right>&=&\frac{1}{T-\Delta}\int_{t_a}^{t_a +T
-\Delta}\langle f(n_a(t,\Delta))\rangle\mathrm{d}t\\
\nonumber
&\sim&C+\frac{1}{T}\int_{t_a}^{t_a +T}\frac{(t)^{\alpha-1}}{\Gamma(\alpha)}
g(\Delta)\mathrm{d}t\\
&=&C+\frac{\Lambda_\alpha(t_a/T)}{\Gamma(1+\alpha)}\frac{g(\Delta)}{T^{1-\alpha}}.
\label{ctrw:eata}
\end{eqnarray}
\end{subequations}
The approximation in the second line holds for sufficiently long trajectories, 
$T\gg\Delta$. Notice that from the full, possibly complicated time dependence of 
the ensemble average, only the late-age limiting behavior, 
Eq.~(\ref{ctrw:ea_high}), entered this approximation. The reason is that with 
the integrand in the time average of Eq.~(\ref{ctrw:eata}) decaying like 
$(t')^{\alpha-1}$, the integral itself is still an increasing function of $T$, 
namely it grows like $T^\alpha$.

Thus, indeed aging effects for time averages in CTRW type of random motion are 
universally described in terms of the constant $C$ and simple prefactors 
$\Lambda_\alpha$ and $T^{\alpha-1}$. The full lag time dependence is captured by 
the function $g(\Delta)$, which is independent of the parameters defining the 
time window of observation, $t_a$ and $T$. Conversely, the concrete choice for 
the model $x(n)$ or the observable $F$ enter only $C$ and $g$, but not the 
prefactors.

Furthermore, the $\Delta$-dependence of the time average is closely related to 
the $t$-dependence of the ensemble average at high ages, $t_a\gg t$. In 
particular, we find a universal asymptotic equivalence in the time scaling of 
time and ensemble averages, if both are taken during late measurements:
\begin{equation}
\left<\overline{F(\Delta;t_a,T)}\right>\sim \langle 
F(x(t_a+\Delta),x(t_a))\rangle \text{ for } t_a\gg T\gg \Delta .
\end{equation}
Thus indeed, averaging at late ages appears to happen under stationary 
conditions. Keep in mind however, that above identity refers to the expectation 
value of time averages, and thus to the time scaling behavior. Since CTRWs exhibit weak ergodicity breaking, the amplitude of the time 
average of a single trajectory can largely deviate from the expected value, 
especially at high ages, See for instance the discussion on the ergodicity 
breaking parameter for $\overline{\delta^2}$ in~\cite{Schulz2013}.

\subsection{Interplay of aging and internal relaxation}\label{ctrw:flectrw}
To conclude the section on aging CTRWs, we study a process which extends the
previous examples to a more elaborate model. On the one hand, this serves as exemplary application of the formulae and methods described in this paper. In particular, it gives a concrete meaning to the analytical discussion of time and ensemble averages and their relation to internal relaxation time 
scales. On the other hand, its complexity and versatility makes it more 
suitable for possibly describing real world physical phenomena (examples below). 
The definition of the model is as follows. 

With our base random walk $x(n)$ we 
depart from the simple Brownian motion and instead consider the
fractional Langevin equation (FLE)
\begin{equation}
 0 =-\lambda x(n) -\overline{\gamma}\int_0^n(n-n')^{2H-2} \,\dot x(n') \,\mathrm{d}n' +\sqrt{\overline{\gamma} k_B\mathcal{T}}\,\xi_H(n) .
 \label{ctrw:fle}
\end{equation}
Here, the dot signals a first order derivative, $\overline{\gamma}>0$ is a generalized friction constant, $k_B\mathcal{T}>0$ gives the thermal energy of the environment, and $\lambda>0$ quantifies the strength of an external, harmonic potential $V(x)=\lambda x^2$, centered around $x=0$. Thus, this 
FLE describes the random motion $x(n)$ of a point-like 
particle subject to the static external potential, a friction force, and a fluctuating force due to the interaction with the surrounding heat bath. We assume that dynamics are overdamped, meaning that the particle mass is so small that we can neglect particle inertia. Consequently, there is no term proportional to $\ddot{x}(n)$ in Eq.~(\ref{ctrw:fle}). The random force $\xi_H$ is modeled in 
terms of the so called fractional Gaussian noise, i.e. a Gaussian 
process defined through \cite{Mandelbrot1968,Gripenberg1996}
\begin{subequations}
\begin{equation}
\langle \xi_H(n)\rangle=0
\end{equation}
and
\begin{eqnarray}
\langle\xi_H(n_1)\xi_H(n_2)\rangle&=&|n_2-n_1|^{2H-2}\nonumber\\
&&\hspace*{-1.8cm}+\frac{2}{2H-1}|n_2-n_1|^{2H-1}\delta(n_2-n_1),
\end{eqnarray}
\label{ctrw:fgn}%
\end{subequations}
where $1/2<H<1$.

While the noise process itself can be defined, in principle, for any $0<H<1$, the friction kernel in Eq.~(\ref{ctrw:fle}) diverges for values $H\leq1/2$. Hence, we restrict ourselves to $1/2<H<1$, implying that noise correlations are of the long-range, persistent type. They stand for the interaction with a complex environment, where relaxation dynamics are slow and cannot be characterized in terms of single relaxation rates. The latter anomalous diffusion phenomenon comes about when accounting for obstructed motion (e.g. single-file diffusion~\cite{Sander2012} or other many-body systems \cite{Lizana2010,Kupferman2004}) or an interaction with a viscoelastic network~\cite{Goychuk2012,Weber2010}. Biological cells feature highly complex, crowded environments, crossed by filament networks. FLE dynamics have thus found wide application in biological physics, describing diffusive motion processes within the cell~\cite{Weber2010,Burnecki2012,Szymanski2009,Jeon2012a}, but also conformational dynamics of individual protein complexes~\cite{Min2005}.

Such long-range correlations carry over to the particle motion in a two-fold way, as also the friction term in Eq.~(\ref{ctrw:fle}) incorporates long-range memory effects:  its magnitude is defined by the complete velocity history starting from  $n'=0$. The power-law memory kernel is chosen such that this FLE fulfills a Kubo-style fluctuation dissipation theorem~\cite{Kubo1966}. This implies that  deterministic friction and random noise forces do not originate from separate physical mechanisms, but are both generated by interactions with the 
environment. Moreover, it means we can model `equilibrated diffusion': 
Eq.~(\ref{ctrw:fle}) admits a solution with a stationary velocity profile, so 
that the net energy exchange of the particle with the heat bath is zero. Such a
solution is a Gaussian process defined by~\cite{Pottier2003}
\begin{subequations}
\begin{equation}
\langle x(n)\rangle=0
\end{equation}
and
\begin{equation}
\langle [x(k+n)-x(k)]^2\rangle=\frac{2k_B\mathcal{T}}{\lambda}\left[1-\mle{2-2H}{
-\frac{\lambda}{\gamma}|n|^{2-2H}}\right],
\end{equation}
\label{ctrw:fle_stat}%
\end{subequations}
in terms of a generalized Mittag-Leffler function $E_{2-2H}$ (appendix~\ref{app:functions}), introducing $\gamma=\overline{\gamma}\Gamma(2H-1)$.
(Note that the definition of a Gaussian process in terms of its correlation 
function is equivalent to the definition in terms of squared increments, since 
$\langle [x_2-x_1]^2 \rangle = \langle x_1^2 \rangle +\langle x_1^2 \rangle 
-2\langle x_1x_2 \rangle$.) The nonequilibrium solutions to the FLE are discussed in~\cite{Deng2009,Kursawe2013}, including an interesting discussion on their transient aging behavior~\cite{Kursawe2013}. In the limit $H\rightarrow1/2$, Eq.~(\ref{ctrw:fle_stat}) defines a stationary Ornstein-Uhlenbeck process, implying exponential relaxation. Physically, the latter models overdamped motion in an harmonic potential where friction and noise forces are memoryless.

The stationary Gaussian process~(\ref{ctrw:fle_stat}) is the starting point for our discussion on aging induced by heavy-tailed waiting times. There are, of course, various ways to introduce an aging, nonergodic, CTRW-like model component, and the resulting stochastic processes differ largely. For instance, one can combine the Gaussian dynamics with an independent CTRW motion in a purely additive manner, as discussed in~\cite{Jeon2013a}. Moreover, for a nonoverdamped, inertial motion, the FLE velocity process $v(n)=\dot{x}(n)$ can be modified by adding periods of constant velocity with heavy-tailed statistics \cite{Friedrich2006}.   Here, we instead stick to the standard subordination approach as described in the preceding sections: we introduce stalling dynamics by defining $x(t)=x(n(t))$, where the aging renewal process $n(t)$ is assumed to be statistically independent from $x(n)$. 
Physically, this scenario implies that the particle under observation is governed
by the FLE (\ref{ctrw:fle}). Eventually, it becomes trapped for a random waiting
time controlled by the probability density $\psi(t)$. After release from the trap
we assume that the particle motion quickly thermalizes and the particle again
follows the stationary Gaussian dynamics (\ref{ctrw:fle_stat}) until the next trapping event.

We basically have three motivations to study this random motion. First, CTRWs provide one approach to model diffusive motion in biological cells, where waiting times represent multi-scale binding or caging dynamics. The sheer complexity of this kind of environment however brings the necessity to further introduce aspects of anomalous diffusion not contained in the renewal process $n(t)$ ~\cite{cells,Meroz2010,Akimoto2011,Weiss2013,Tabei2013,Hoefling2013}. While the overdamped FLE dynamics~(\ref{ctrw:fle_stat}) adds both friction and external binding forces, it also introduces the concept of long-time correlations within an equilibrated environment. 

From the point of view of our theoretical 
discussion of aging CTRWs, our second motive to discuss this particular model is 
the stationarity of increments of $x(n)$. It allows us to exercise the methods 
we introduced in the previous section by calculating ensemble and time
averages---mean squared displacements in particular---of the aging process $x(t)$. 

Third, aside from its didactic purpose, the definition of the process $x(n)$ also 
extends the ordinary Brownian motion by introduction of an intrinsic time scale 
$n^*=(\gamma/\lambda)^{1/(2-2H)}$, characterizing the competition between binding and friction forces. According to Eq.~(\ref{ctrw:fle_stat}), stationary Langevin dynamics exhibit a turnover
\begin{align}
 &\langle [x(k+n)-x(k)]^2 \rangle \sim \frac{2k_B\mathcal{T}}{\lambda}\nonumber\\
 &\qquad\times\begin{cases}
   [\Gamma(3-2H)]^{-1} (n/n^*)^{2-2H}, &n\ll n^*\\[0.2cm]
   1, &n\gg n^*
 \end{cases},
\end{align}
from subdiffusion on short time scales to the stationary Boltzmann-limit $2k_B\mathcal{T}/\lambda$ on long scales. Now for the aging process $x(t)$, we know that the age of a measurement $t_a$ itself can be conceived as a time scale separating the time dependence into early and late age behavior. It would be interesting to know how complex the process $x(t)$ behaves with respect to both time scales. Are all conceivable time regimes clearly distinct? Are time and ensemble averages equally sensitive to transitions from one regime to the other?

A partial answer is given by Eqs.~(\ref{ctrw:ea_highA}), which 
focus on the calculation of ensemble averages. We discuss the mean squared 
displacement, $F(x_1,x_2)=(x_1-x_2)^2$ and $f(n)$ as defined through 
Eqs.~(\ref{ctrw:fle_stat}). Inserting the latter into Eq.~(\ref{ctrw:ea_low}), we find an approximation for low ages, $t_a\ll t$:
\begin{eqnarray}
&&\langle x_a^2(t_a,t)\rangle=\langle [x(t_a+t)-x(t_a)]^2\rangle\nonumber\\
&&
\begin{aligned}
&=\langle [x(t)-x(0)]^2\rangle + \mathcal{O}((t_a/t)^\alpha) \\ 
&\sim\frac{2k_B\mathcal{T}}{\lambda}\mathscr{L}^{-1}_{s\to t/\tau}\Big\{s^{
\alpha-1}\\
    &\quad\times\mathscr{L}_{n\to s^{\alpha}}\left\{\left[1-E_{2-2H}\left(-\frac{
\lambda}{\gamma}|n|^{2-2H}\right)\right]\right\}\Big\}\\
    &= \frac{2k_B\mathcal{T}}{\lambda} \,\Lti{\frac{1}{s}-\frac{s^{(2-2H)\alpha-1}}{s^{(2-2H)\alpha}-\lambda/\gamma)}}{s}{t/\tau} \\
    &= \frac{2k_B\mathcal{T}}{\lambda} \,\left[1-\gmle{(2-2H)\alpha}{2-\alpha}{-\frac{\lambda}{\gamma_\alpha} t^{(2-2H)\alpha}}\right] .
 \end{aligned}
 \label{ctrw:fle_low}\nonumber\\
\end{eqnarray}
In order to obtain reasonable physical units, we reintroduced the parameter 
$\tau$ from Eq.~(\ref{intro:tail}) bearing the dimension of seconds. 
Moreover, the new constant $\gamma_\alpha=\gamma\tau^{(2-2H)\alpha}$ has physical units $\text{g s}^{-(2-2H)\alpha-2}$. 

For increasingly aged ensemble measurements, aging corrections of the order 
$(t_a/t)^\alpha$ come into play. The detailed time behavior of the ensemble mean squared displacement is found by combining Eqs.~(\ref{ctrw:fle_stat}), (\ref{ctrw:ea}) and~(\ref{renewal:ameanf}). We here have $f(0)=0$, and thus we can write
\begin{subequations}
\begin{eqnarray}
 && \langle x_a^2(t_a,t)\rangle =  \frac{2k_B\mathcal{T}}{\lambda} \times \nonumber\\
 &&\quad h(t_a,t) *_t \left[1-\gmle{(2-2H)\alpha}{2-\alpha}{-\frac{\lambda}{\gamma_\alpha} t^{(2-2H)\alpha}}\right]. \nonumber\\
 \label{ctrw:fle_complete}
\end{eqnarray}
The full time dependence of the ensemble average comes as a convolution of the forward recurrence time PDF~(\ref{renewal:rec}) with a generalized Mittag-Leffler function. We provide graphical examples below.

The behavior at high ages, $t_a\gg t$ can be calculated analytically by use of Eqs.~(\ref{ctrw:fle_stat}) and~(\ref{ctrw:ea_low}), yielding
\begin{equation}
 \langle x_a^2(t_a,t) \rangle \sim \frac{t_a^{\alpha-1}}{\Gamma(\alpha)}\,g(t) ,
\end{equation}
where in this case $C=f(0)=0$ and
\begin{eqnarray}
\nonumber
g(t)&=&\frac{2k_B\mathcal{T}}{\lambda}\mathscr{L}^{-1}_{s\to t/\tau}\Big\{s^{2
\alpha-2}\\
\nonumber
&&\hspace*{0.8cm}\times\Lt{1-\mle{2-2H}{-\frac{\lambda}{\gamma}|n|^{2-2H}}}{n}{s^
\alpha}\Big\}\\
\nonumber
&=&\frac{2k_B\mathcal{T}}{\lambda}t^{1-\alpha}\left[\frac{1}{\Gamma(2-\alpha)}
\right.\\
&&\left.-\gmle{(2-2H)\alpha}{2-\alpha}{-\frac{\lambda}{\gamma_\alpha}t^{(2-2H)
\alpha}}\right]. 
\label{ctrw:fle_high}
\end{eqnarray}
\end{subequations}
The time dependence of the mean squared displacement, Eq.~(\ref{ctrw:fle_complete}), is relatively complicated. 
In both the limits of slightly and highly aged measurements we observe a 
Mittag-Leffler behavior, but with different parameters. We can extract from 
Eqs.~(\ref{ctrw:fle_low}) and~(\ref{ctrw:fle_high}) four time regimes where the 
diffusive behavior with respect to time $t$ is described in terms of single 
power-laws; regimes are separated by a time scale $t_a$ induced by aging and an 
intrinsic relaxation time scale $\tau_\alpha^*=(\gamma_\alpha/\lambda)^{1/[(2-2H)\alpha]}$,
\begin{subequations}
\begin{equation}
 \langle x_a^2(t_a,t) \rangle \sim\frac{2k_B\mathcal{T}}{\lambda}
 \begin{cases}
  A_\alpha^*\, t_a^{\alpha-1}t^{1-(2H-1)\alpha} , &t\ll
t_a,\tau_\alpha^*, \\[0.2cm]
  B_\alpha^*\, t^{(2-2H)\alpha} , &t_a\ll t\ll\tau_\alpha^*, \\[0.2cm]
  C_\alpha^*\, t_a^{\alpha-1}t^{1-\alpha} , &\tau_\alpha^*\ll t\ll 
t_a, \\[0.2cm]
  1 , &t_a,\tau_\alpha^*\ll t,
 \end{cases}
 \label{ctrw:fle_regimes}
\end{equation}
with coefficients
\begin{align}
\nonumber
 A_\alpha^*&= \left[{\tau_\alpha^*}^{(2-2H)\alpha}\Gamma(\alpha)\Gamma(2-(2H-1)\alpha)\right]^{-1} ,\\
\nonumber
 B_\alpha^*&= \left[{\tau_\alpha^*}^{(2-2H)\alpha}\Gamma(1-(2-2H)\alpha)\right]^{-1} ,\\
 C_\alpha^*&= \left[\Gamma(\alpha)\Gamma(2-\alpha)\right]^{-1} .
\end{align}
\end{subequations}

\begin{figure}
 \includegraphics{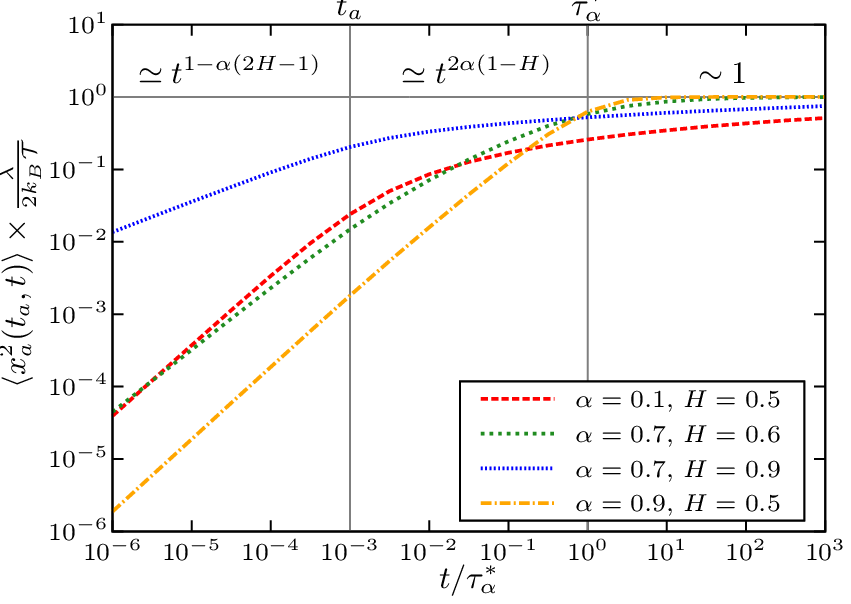}
 \includegraphics{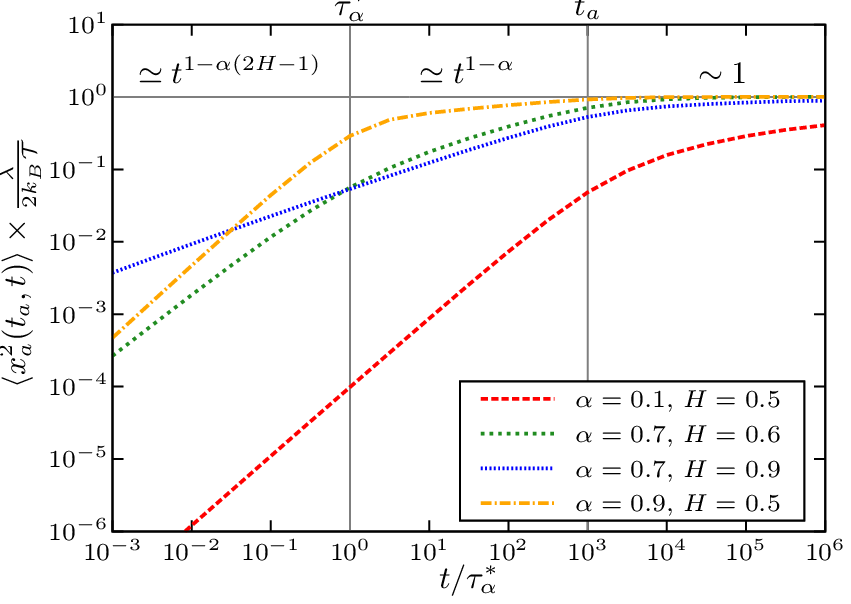}
 \caption{
 Time evolution of the ensemble averaged mean squared displacement for 
combined CTRW and overdamped, confined FLE dynamics. Plots are numerical evaluations of the convolution integral~(\ref{ctrw:fle_complete}). We study several parameter configurations for $H$ and $\alpha$. The behavior is characterized by a double turnover between several power-law regimes as labeled above the graphs, see also Eq.~(\ref{ctrw:fle_regimes}). Respective turnover time scales $\tau_\alpha^*$ and $t_a$ are indicated as vertical lines. A horizontal line gives the infinite time stationary limit $\langle x_a^2\rangle\rightarrow 2k_b\mathcal{T}/\lambda$. \textit{Top:} The ensemble measurement starts at a 
time where internal FLE dynamics have not yet relaxed to equilibrium, i.e. 
$t_a\ll\tau^*_\alpha$. \textit{Bottom:} Opposite case, $t_a\gg\tau^*_\alpha$. }
 \label{fig.fle_ea}
\end{figure}

Figure~\ref{fig.fle_ea} gives several examples for the detailed turnover 
behavior of the mean squared displacement for various values of $\alpha$ and 
$H$. At infinite times $t\rightarrow\infty$, the ensemble mean squared displacement converges to the Boltzmann limit. At finite times, we generally observe subdiffusive behavior. The dynamics are slowest when $t$ is short as compared both intrinsic relaxation and aging time scales. Notice that the behavior at times $t\gg\tau_\alpha^*$ is independent of the parameter $H$ defining the memory of friction and noise forces. This regime is fully dominated by the aging transition.

Again we stress that the full double turnover behavior of the ensemble average 
might not be visible to the observer of a real physical system due to 
limitations of the experimental setup. In addition, precise knowledge on the 
aging time $t_a$, which preceded the actual ensemble measurement, might not be 
available. Maybe $t_a$ is even random, varying from one trajectory to the next. 
In any such case, the observer cannot know which of the power-law regimes 
of~(\ref{ctrw:fle_regimes}) he is probing by means of mean squared displacement 
measurements.

The analog time average measurement is much less complex and thus easier to 
interpret. According to Eq.~(\ref{ctrw:eata}) we have for $T\gg\Delta$,
\begin{eqnarray}
 &&\left<\overline{\delta^2(\Delta;t_a,T)}\right>\sim \frac{\Lambda_\alpha(t_a / 
T)}{\Gamma(1+\alpha)} \; \frac{g(\Delta)}{T^{1-\alpha}} \nonumber\\
\nonumber
 &&= \frac{\Lambda_\alpha(t_a / T)}{\Gamma(1+\alpha)\,T^{1-\alpha}} \; 
\frac{2k_B\mathcal{T}}{\lambda}\Delta^{1-\alpha}\left[\frac{1}{\Gamma(2-\alpha)}
\right.\\
&&\left.\quad-\gmle{(2-2H)\alpha}{2-\alpha}{-\frac{\lambda}{\gamma_\alpha}
\Delta^{(2-2H)\alpha}}\right].
 \label{ctrw:fle_eata}
\end{eqnarray}
The dependence on age and measurement time parameters, $t_a$ and $T$ factorizes from the 
lag time dependence. The latter is captured by the function $g$. Recall that for 
weakly nonergodic CTRW dynamics, the amplitude of a single-trajectory time 
average $\overline{\delta^2}$ is random by nature, while its scaling with lag 
time $\Delta$ is not. Thus, the $\Delta$-scaling of the time averaged mean 
squared displacement is age-independent. For combined CTRW-FLE dynamics, it is 
universally given by $g$ in terms of a Mittag-Leffler type single turnover, with 
(lag) time regimes being separated by the intrinsic time scale 
$\tau^*_\alpha=(\gamma_\alpha/\lambda)^{1/[(2-2H)\alpha]}$. In particular, a single, long 
trajectory measurement is in principle sufficient to read off the scaling 
exponents $\alpha$ and $H$, and the time scale $\tau^*_\alpha$.

\begin{figure}
 \includegraphics{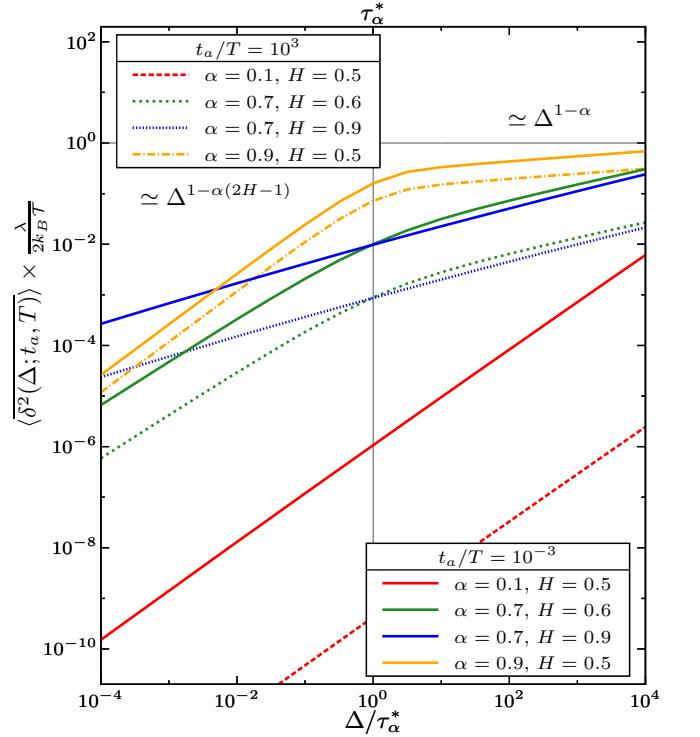}
 \caption{Expectation value of the time averaged squared displacement 
$\overline{\delta^2}$ as a function of lag time $\Delta$. Plots graphically 
represent Eq.~(\ref{ctrw:fle_eata}), for the same system parameters $\alpha$ and $H$ as in Fig.~\ref{fig.fle_ea}. In all cases the duration of the measurement is $T=10^6$. In contrast to the analog ensemble average, the behavior is given by a single turnover at the internal relaxation time scale $\tau^*_\alpha$, no matter at which time $t_a$ the recording of the trajectory was initiated. Results obtained at late age ($t_a/T=10^3$, dashed/dotted lines) differ from those at early age ($t_a/T=10^{-3}$, continuous lines of same color) merely by a prefactor 
(displaying as a constant shift in the double logarithmic plot).}
\label{fig.fle_eata}
\end{figure}

Aging affects the amplitude of the time average only: as $t_a$ increases, we 
expect smaller values of $\overline{\delta^2}$. For exemplary plots, see 
Fig.~\ref{fig.fle_eata}. Notice that the late lag time behavior is generally $\Delta^{1-\alpha}$, independently of $H$, as reported previously~\cite{Neusius2009,Burov2010} for confined, memoryless CTRWs (i.e. for $H=1/2$). Moreover, in the limit $\alpha\rightarrow1$, aging becomes negligible, $\Lambda_1\equiv1$, and we recover the ergodic FLE result $\langle\overline{\delta^2(\Delta)}\rangle=\langle[x(n=\Delta/\tau)-x(0)]^2\rangle$, with $x(n)$ as in Eqs.~(\ref{ctrw:fle_stat}).

\section{Conclusions}
We investigated a renewal process in which the average waiting time between 
consecutive renewals is infinite. For such a process, randomness is relevant on 
even the longest of time scales. We worked out several nontrivial properties, and
here we concentrated on the phenomenon of
aging: the statistics of renewals counted within a finite observation 
time window strongly depends on the specific instant at which we start to count.
The most remarkable aging effect is a growing, discrete probability to 
not count a single event during observation.
Concurrently, also the continuous distribution of the nonzero number of renewals is 
transformed. We discussed analytical expressions for this distribution with 
respect to series expansions, tail behavior and monotonicity, see Eqs.~(\ref{renewal:aPDF_early_LaplaceA}) to (\ref{renewal:aPDF_high_FoxHA}) and Figs.~\ref{fig.malpha} and~\ref{fig.PDF}. We deduced exact 
expressions for related ensemble averages, Eqs.~(\ref{renewal:ameanf_limA}), such as arbitrary moments of the number of renewals in terms of incomplete Beta functions, Eq.~(\ref{renewal:nq}). We found that we count fewer renewals at increasingly late 
ages. However, this is mainly due to the high probability of counting zero 
renewals. Restricting the counting statistics to eventful measurements, see Eq.~(\ref{renewal:nq_conditional}), yields a finite distribution even at infinite age, characterized by the same time scaling as the non-aged process.

We applied this aging renewal theory to CTRW models: 
renewal events are identified with steps of the random walk. The
aging effects translate from the renewal to the random walk process. Thus the 
increasing probability for the complete absence of dynamic activity is conceived as a population splitting effect. In an ensemble of identical 
particles, a certain fraction remains fully immobile during a finite time 
observation. The total size of the mobile population decreases with age, but 
also their detailed statistics change.

We discussed the implications for ensemble and time averaged observables in CTRW theory. Aging affects the distributions of time averages, and population 
splitting has to be considered in particular. Remarkably, we find that ensemble averages behave very differently with 
respect to aging effects than the analogous time averages. The age of a measurement 
modifies the complete time dependence of an ensemble average, mimicking an 
internal relaxation mechanism. We provide an analytical description in terms of Eqs.~(\ref{ctrw:ea_highA}). In contrast, aging enters the associate time average only as a distributional modification, while its scaling with respect to (lag) time remains indifferent. We calculated the respective scaling function $g$ [see Eqs.~(\ref{ctrw:ea_high}) and~(\ref{ctrw:eata})]. In addition, we derived the precise aging modifications for the ensemble averaged time average. The latter in turn do not depend on details of the definition of the process or the observable. Instead they are captured by universal constants, and the age enters in particular through the aging depression $\Lambda_\alpha$, defined in Eq.~(\ref{ctrw:depression}). Despite this fundamental conceptual difference between time and ensemble averages, we found that their time scalings are identical in highly aged measurements. This 
is a surprising result since CTRWs are notorious for weak ergodicity 
breaking, i.e. the general \emph{in}equivalence of the two types of averages.
 
We bestowed a more specific meaning to these general considerations by discussing 
combined FLE-CTRW dynamics. The latter extend ordinary CTRW models
by introduction of binding and friction forces and a correlated noise. We contrasted turnover behaviors of aging and internal relaxation, and gave a detailed discussion of associated limiting regimes.

The CTRW is a very natural application of aging renewal theory, yet it is far 
from unique. All aging effects such as population splitting and altered ergodic 
behavior have their analog phenomenon in any physical system where the mean 
sojourn time in microstates is infinite. Thus we expect a certain fraction of blinking quantum dots, or cool atoms, to be constantly stuck in one state during a delayed observation period. At the same time the statistics of the switching population are aging. Power spectral densities obtained from signals from such systems should display related aging properties, as discussed briefly in~\cite{Niemann2013}. Aside from stochastic process studies, also weakly chaotic systems were shown to exhibit analogous aging behaviors~\cite{Akimoto2013}. When it comes to diffusion dynamics, a study of alternative CTRW-like models might turn out worthwhile. Examples are the noisy CTRW~\cite{Jeon2013a}, aging renewals on a velocity level~\cite{Friedrich2006}, L{\'e}vy walks~\cite{Shlesinger1987}, non-renewal, correlated CTRW~\cite{Tejedor2010}, and aging CTRW \cite{lomholt}. Moreover, further studies of the aging renewal process, e. g. with respect to higher dimensional probability distributions, will reveal additional insight into aging mechanisms of such physical systems.

\acknowledgments
Johannes Schulz was supported by the Elitenetzwerk Bayern in the framework of the doctorate program Material Science of Complex Interfaces. We acknowledge funding from the Academy of Finland (FiDiPro scheme), and the Israel Science Foundation.

\appendix

\section{Special functions}\label{app:functions}

The asymptotics of the PDF $p_a$ of the aging renewal process $n_a$, 
Eqs.~(\ref{renewal:aPDF_early_FoxH}) and~(\ref{renewal:aPDF_high_FoxH}), and the rescaled 
TAMSD distribution in Eq.~(\ref{ctrw:tamsd_pdf}) for highly aged processes 
contain the Fox $H$-function, a very convenient special function \cite{Mathai}.
For the specific cases considered here the series expansion around $z=0$ is
\begin{equation}
\FH{1}{0}{1}{1}{z}{(\beta,\alpha)}{(0,1)}
=\sum_{k=0}^\infty\frac{(-z)^k}{k!\Gamma(\beta-\alpha k)} .
\end{equation}
The behavior for large values of $z$ is stretched exponential,
\begin{equation}
\FH{1}{0}{1}{1}{z}{(\beta,\alpha)}{(0,1)}
\sim Cz^{\frac{1-2\beta}{2-2\alpha}}\exp(-Dz^{\frac{1}{1-\alpha}}),
\end{equation}
with the abbreviations
\begin{equation}
C=[2\pi(1-\alpha)]^{-\frac{1}{2}}\alpha^{\frac{1-2\beta}{2-2\alpha}},\qquad
D=(1-\alpha)\alpha^{\frac{\alpha}{1-\alpha}}.
\end{equation}
Logarithmic tail analysis thus yields
\begin{equation}
\log\left\{\FH{1}{0}{1}{1}{z}{(\beta,\alpha)}{(0,1)}\right\}
\simeq -z^{\frac{1}{1-\alpha}} ,
\end{equation}
independently of $\beta$. The expressions for $p_a$ as non-aged, slightly aged 
and strongly aged renewal PDF are obtained by substituting $\beta$ with 
$1-\alpha$, $1-2\alpha$ and $2-2\alpha$, respectively. Note that for 
$\alpha=1/2$ the relevant $H$-functions simply become
\begin{eqnarray}
  \FH{1}{0}{1}{1}{z}{(1/2,1/2)}{(0,1)}
  &=& \frac{1}{\sqrt{\pi}}\, \exp(-z^2/4) ,\nonumber\\
  \FH{1}{0}{1}{1}{z}{(0,1/2)}{(0,1)}
  &=& \frac{z}{\sqrt{4\pi}}\, \exp(-z^2/4) ,\nonumber\\
  \FH{1}{0}{1}{1}{z}{(1,1/2)}{(0,1)}
  &=& \text{erfc}(z/2)
\end{eqnarray}
in terms of exponential and complementary error functions.

The probability $m_{\alpha}$ in Eq.~(\ref{renewal:aPDF}) and the $q$th order 
moments of renewals~(\ref{renewal:nq}) are expressed in terms of an incomplete 
Beta function, defined through \cite{Abramowitz}
\begin{eqnarray}
B(z;a,b)&=&\int_0^z u^{a-1}(1-u)^{b-1}\,du\nonumber\\[0.2cm]
 &\sim&\left\{\begin{array}{ll}z^a/a,\qquad\quad&z\gtrsim0\\[0.2cm]
B(a,b)-z^b/b,
&z\lesssim1\end{array}\right.,
\end{eqnarray}
with the special value
\begin{equation}
 B(1;a,b)=B(a,b)=\Gamma(a)\Gamma(b)/\Gamma(a+b).
\end{equation}
Here $a,b>0$ and $0\leq z\leq1$.

The Laplace transform of $p_a$ with respect to the number of events, 
Eqs.~(\ref{renewal:nexp}) and~(\ref{renewal:nexp_lim}), and the mean squared 
displace for FLE motion, Eqs~(\ref{ctrw:fle_stat}), (\ref{ctrw:fle_low}) 
and~(\ref{ctrw:fle_high}) are expressed in terms of generalized Mittag-Leffler 
functions. The latter are characterized alternatively by series expansions 
around $z=0$,
\begin{equation}
  E_{\alpha,\beta}(z) = \sum_{k=0}^{\infty} \frac{z^k}{\Gamma(\alpha k+\beta)},
\end{equation}
 asymptotic series for large arguments,
\begin{equation}
  E_{\alpha,\beta}(z) \sim -\sum_{k=1}^\infty \frac{z^{-k}}{\Gamma(\beta-\alpha 
k)},
\end{equation}
or by their Laplace pair
\begin{equation}
  E_{\alpha,\beta}\left(-zt^\alpha\right) = t^{1-\beta} 
\,\Lti{\frac{s^{\alpha-\beta}}{s^\alpha+z}}{s}{t}
\end{equation}
for any $\alpha,\beta>0$. The ordinary Mittag-Leffler functions are the special 
cases $\mle{\alpha}{z}\equiv\gmle{\alpha}{1}{z}$.


\begin{thebibliography}{}
\bibitem{footnote:plotstable}
Numerical tools for computing stable densities are available for common programs such as Mathematica or Matlab.
\bibitem{footnote:laplacetrafo}
We express the Laplace transform $f(s)=\Lt{f(t)}{t}{s}=\int_0^\infty f(t) \exp(-st) \,\mathrm{d}t$ of a function $f(t)$ by explicit dependence on the Laplace variable $s$. Likewise, $f(s_a)$ and $f(t_a)$ are Laplace pairs, and Laplace inversion is occasionally indicated explicitely as $f(t)=\Lti{f(s)}{s}{t}$.
\end{thebibliography}
\end{document}